\documentclass{aa}
\usepackage{amsmath,graphicx,amssymb}
\usepackage{txfonts}
\usepackage{natbib}
\usepackage{xcolor}
\usepackage[T1]{fontenc}
\usepackage{url}
\usepackage{lipsum}
\definecolor{DarkBlue}{rgb}{0.0,0.1,0.4}
\definecolor{Red}{rgb}{0.6,0.2,0.1}
\definecolor{DarkRed}{rgb}{0.5,0.0,0.5}
\definecolor{Blue}{rgb}{0.0,0.0,1.0}
\definecolor{Zelinkava}{rgb}{0.2,0.5,0.2}
\definecolor{Pink}{rgb}{0.0,0.7,0.7}
\definecolor{White}{rgb}{1.0,1.0,1.0}
\usepackage{hyperref}
\bibpunct{(}{)}{;}{a}{}{,}

\hypersetup{
    colorlinks=true,
    linkcolor=Blue,
    citecolor=Blue,
    filecolor=Blue,     
    urlcolor=Blue,
}

\renewcommand{\d}{\text{d}}
\newcommand{\dd}{\,\text{d}}

\begin{document}

\title{Modeling of interactions between supernovae ejecta and  aspherical circumstellar environments}

\author{P.~Kurf\" urst\inst{1,2} \and J.~Krti\v{c}ka\inst{1}}

\institute{Department of Theoretical Physics and Astrophysics,
           Masaryk University, Kotl\' a\v rsk\' a 2, CZ-611\,37 Brno, Czech Republic\and
           Institute of Theoretical Physics, Charles University, V Hole\v sovi\v ck\' ach 2, 180 00 Praha 8, Czech Republic}

\date{Received}

\abstract {Massive stars are characterized by a significant loss of mass 
either via (nearly) spherically symmetric stellar winds or pre-explosion pulses,
or by aspherical forms of circumstellar matter (CSM) such
as bipolar lobes or outflowing circumstellar equatorial disks. 
Since a significant fraction of most massive stars 
end their lives by a core collapse, supernovae (SNe) are always
located inside large circumstellar envelopes created by their progenitors.} 
{We study the dynamics and thermal effects of collision between expanding 
ejecta of SNe and CSM that may be formed during,
for example, a sgB[e] star phase, a luminous blue variable phase, around
PopIII stars, or by various forms of accretion.}
{For time-dependent hydrodynamic modeling we used our own grid-based Eulerian multidimensional
hydrodynamic code built with a finite volumes method. 
The code is based on a directionally unsplit Roe's method that is highly efficient for 
calculations of shocks and physical flows with large discontinuities.}
{We simulate a SNe explosion as a spherically symmetric blast wave. 
The initial geometry of the disks corresponds to a density structure of a material that orbits in Keplerian trajectories.
We examine the behavior of basic hydrodynamic characteristics, i.e., the
density, pressure, velocity of expansion,
and temperature structure in the interaction zone under various 
geometrical configurations and various initial densities of CSM.
We calculate the evolution of the SN - CSM system and 
the rate of aspherical deceleration as well as the degree
of anisotropy in density, pressure, and temperature distribution.}
{Our simulations reveal significant asphericity
of the expanding envelope above all in the case of 
dense equatorial disks.
Our ``low density'' model however also shows significant asphericity in the case of the disk mass-loss rate 
$\dot{M}_\text{csd}=10^{-6}\,M_\odot\,\text{yr}^{-1}$.
The models also show the zones of overdensity in the SN - disk contact region and indicate the 
development of Kelvin-Helmholtz instabilities within the zones of shear 
between the disk and the more freely expanding material outside the disk.}


\titlerunning{Modeling of adiabatic interactions between supernovae ejecta and circumstellar
disks}

\authorrunning{P.~Kurf\"urst et al.}
\maketitle

\section{Introduction}

Explosions of supernovae (SNe) are one of the most prominent events in the universe. 
They may provide excellent observational and theoretical material for studying a vast range of 
physics: from cosmological principles to particle physics. The interaction of the SN blast wave may serve as a
probe of the progenitor mass loss and its circumstellar matter (CSM).

Most of the core-collapse (type II) SN progenitors are considered to be red supergiants 
\citep{2009ARA&A..47...63S,2015PASA...32...16S,2015ApJ...806..225P,2017ApJ...841..127M}, 
however, there is strong evidence that
blue supergiants (BSG) can also be hydrogen-rich massive progenitors of type II SN events 
\citep[e.g.,][]{2013A&A...552A.105V}.
Most famous example is SN 1987A, which in the time of explosion 
was a BSG star evolved from the previous stage of red supergiant; see, for example, 
\citet{2015A&A...581A..36M}.
Furthermore, there are indications of a strong connection between 
BSG progenitors and superluminous SNe that are likely associated with 
very massive stars that retained their thick hydrogen envelopes and that 
explode within the dense CSM environment. There 
is also a strong evidence of rugged and anisotropic nature of the CSM, 
since some of these stars may have been similar to luminous blue variables
that lost a large amount of mass, up to several $M_\odot$, prior to explosion 
\citep{2007ApJ...666.1116S,2008ApJ...686..467S,2013Natur.494...65O,2017hsn..book..403S}. 

The interaction of SNe with CSM may significantly power and modify the observed luminosity.
After a radiation mediated shock goes through the stellar body of a SN progenitor, then 
it continues propagating through CSM. 
Depending on the optical thickness of CSM in the direction to the observer, the radiation 
is in the first phase more or less absorbed 
while an increasing amount of energy is accumulated in the shocked region \citep{2012ApJ...759..108S}.
After the optical depth of CSM drops, this energy is released as a breakout pulse that is much more energetic 
than a CSM-less breakout, however, 
the energy release in the presence of CSM
happens over much longer timescales.
Yet the energy behind the shock grows, if the CSM density does not fall abruptly, after this breakout, 
which is accompanied by the characteristic profile of the light curve 
\citep[e.g.,][]{2011ApJ...729L...6C,2012ApJ...759..108S,2017hsn..book..403S}.

It follows from the nature of (some) BSG stars that during their pre-explosion epochs a CSM may be formed that is anisotropic. In order to make a simulation of a 
hydrodynamic interaction of exploding SN progenitor with such aspherical CSM consisting of 
isotropic stellar wind component and a circumstellar equatorial disk, we consider some typical 
representatives of relevant types of SN progenitors that may be associated with BSG stars and with various types 
of circumstellar disks, such as 
Pop III stars, sgB[e] stars and luminous blue variables 
\citep{1991MNRAS.250..432L,2011A&A...527A..84K,2014A&A...569A..23K}.
We regard the structure of circumstellar disks for the purpose of modeling as 
that corresponding to viscous outflowing disks, which are typical for classical Be stars.
However, we may assume to find this disk type in sgB[e] stars as well 
\citep[for a review see, e.g.,][]{1998ASSL..233....1Z,2006ASPC..355...39H},
where a disk or rings of high density material have been detected \citep{2013A&A...549A..28K}. 
Other types of dense equatorial disks or 
disk-like density enhancements may also be formed owing to, for example, magnetically
compressed winds or binarity and accretion around certain classes of, for instance, B[e] stars 
\citep[e.g.,][]{2006ASPC..355...39H}, luminous blue variables 
\citep[e.g.,][]{1994ApJ...429..846S,2005A&A...439.1107D}, and post-AGB stars 
\citep[e.g.,][]{1998A&A...334..210H}.

Although there have been a number of studies regarding interactions of SNe 
expansion with various forms of spherically symmetric CSM, 
there is a lack of multidimensional hydrodynamic models that describe 
SNe interaction with dense aspherical CSM containing 
equatorial disks. 
The importance of examining such a CSM configuration is further 
enhanced by the fact that the disk densities can be significantly higher in 
the near-stellar region than the 
spherically symmetric wind densities in the case of the same stellar mass-loss rate $\dot M$. 
This is because of their concentration only in the stellar equatorial area along with their much lower velocity of radial outflow.
We calculate time dependent evolution of such SN - CSM interactions for three different values of 
pre-explosion mass-loss rate $\dot{M}$, which led to different masses and densities of CSM and
different ratios of isotropic wind and circumstellar disk densities. 
We use our models to explore and compare the profiles of basic hydrodynamic quantities 
under the significantly aspherical conditions.
We ignore the initial density profile of the progenitor star and the 
effects of radiative cooling and other nonadiabatic processes in the models
because our simulations cover only a relatively short post shock-emergence time.

\section{Basic physics and parameterization of the circumstellar medium}
\label{baseq}
We studied the interaction of a SNe blast wave with the CSM consisting in
the equatorial disk and spherically symmetric CSM (stellar wind).
\subsection{Spherically symmetric CSM}\label{windbaseq}
We assume a spherically symmetric CSM
with a density profile given as a power law
\begin{align}\label{spheresym}
\rho_{\text{sw}}=\rho_\text{0,sw}(R_\star/r)^{w_{\text{sw}}},
\end{align}
where $r$ is radius in spherical coordinates and $R_\star$ is the radius of the central object.
We examined various values of spherically symmetric CSM base density $\rho_\text{0,sw}$ as well as various 
values of spherically symmetric CSM density slope factor $w_{\text{sw}}$ 
within the range from $w_{\text{sw}}=0$ to $w_{\text{sw}}=5$. 
However, the naturally expected structures of the spherically symmetric CSM 
that correspond to a steady mass loss $\dot{M}_\text{sw}=\text{const.}$ fulfill
$\rho_\text{sw}=\dot{M}_\text{sw}/(4\pi r^2\varv_{\text{sw}})\propto r^{-2}$
\citep[][among others]
{1989ApJ...341..867C,1994ApJ...420..268C,2011ApJ...729L...6C,2013MNRAS.428.1020M} 
where we input the average initial outflow velocity
$\varv_\text{sw}=100\,\text{km}\,\text{s}^{-1}$ \citep{2014MNRAS.439.2917M}. 
The adopted spherically symmetric CSM base density ranges from
$\rho_\text{0,sw}\approx10^{-10}\,\text{kg m}^{-3}$ (corresponding to stellar winds of OB stars; see,
e.g.,~\citet{2000ARA&A..38..613K})
to $\rho_\text{0,sw}\approx 10^{-6}\,\text{kg m}^{-3}$, created, for example, 
by enhanced mass loss from violent convective motion caused by the unstable 
nuclear burning at latest stages before core collapse or
by pre-explosion pulses of a progenitor prior to a SN event
\citep[cf.][]{2014MNRAS.439.2917M,2014ApJ...785...82S}.
Since the interaction of the SN ejecta with a spherically symmetric CSM 
is not a key problem in this study, we do not investigate finer details in this point.  

\subsection{Circumstellar equatorial disks structure}\label{circbaseq}
The disks around massive stars result
either from the rotation of contracting matter and from the evacuation of its angular momentum 
(accretion disks, see, e.g., 
\citet{1981ARA&A..19..137P,2002apa..book.....F,2009pfer.book.....M}) 
or they stem from the angular momentum loss from fast rotating central object 
(stellar decretion disks, e.g., 
\citet{1991MNRAS.250..432L,2001PASJ...53..119O,2011A&A...527A..84K,2014A&A...569A..23K}). 

In the following we briefly describe the equations that are essential for modeling the 
structure of a rotating Keplerian disk in this study:
stationary, vertically integrated, and axisymmetric equation of continuity takes the cylindrical form
\begin{align}\label{mascon}
\dot M=2\pi R\Sigma V_R=\text{const.,}
\end{align}
where $\dot M$ is the (constant) disk mass-loss rate, $R$ is
radius in cylindrical coordinates,
$\Sigma=\int_{-\infty}^\infty\rho\dd z$ is the disk surface density
in 1D thin disk approximation, 
and $V_R$ is the velocity of a radial flow of the disk matter.
We hereafter distinguish between cylindrical ($R,\phi,z$)
and spherical ($r,\theta,\phi$) coordinates
using uppercase and lowercase letters, i.e., radius and velocities in
cylindrical coordinates are denoted using uppercase letters, while radius and
velocities in spherical (and Cartesian) coordinates are denoted using lowercase
letters.

Disk density, pressure, and temperature are determined in general by equations 
of radial momentum and angular momentum,
which we present in their stationary form 
\citep[e.g.,][]{2001PASJ...53..119O,2011A&A...527A..84K,2014A&A...569A..23K},
\begin{align}
&V_R\frac{\partial V_R}{\partial R}-
R\Omega^2+\frac{1}{\Sigma}\frac{\partial\,(a^2\Sigma)}
{\partial R}-\mathcal{F}=0,\\
&\frac{\dot M}{2\pi}\frac{\partial\,(R^2\Omega)}{\partial R}-\frac{\partial}{\partial R}
\left(\alpha_\text{vis}\,a^2R^2\Sigma\,\frac{\partial\ln\Omega}{\partial\ln R}\right)=0, 
\end{align}
where $\Omega$ is the angular velocity, $a$ is the speed of sound, and $\alpha_{\text{vis}}$ 
is the parameter of viscosity \citep{1973A&A....24..337S}.
The term $\mathcal{F}=-\Sigma^{-1}\int_{-\infty}^\infty\rho g_R\,\text{d}z$ 
is the density-weighted vertically integrated radial component of gravitational force per unit volume, 
$\rho g_R\equiv\rho GM_\star R/r^3$,
in the thin disk approximation, $z\ll R$, where
$G$ is the gravitational constant and
$M_\star$ is the mass of a central object (star).

We assume a disk initial radial temperature profile for the purpose of the study
in a form of the simple power law
\begin{align}\label{temperature}
T=T_0(R_{\star}/R)^p, 
\end{align}
where $T_0$ is the temperature of the disk near the stellar surface (disk base temperature) and
$p$ is a free parameter ($p\geq 0$) with theoretically
estimated values in range between $0$ and $0.4$ \citep[cf.][]{2008ApJ...684.1374C}.
For simplicity we omit the vertical (relatively shallow) variations of the temperature profile
\citep[e.g.,][]{2007ApJ...668..481S,2018A&A...613A..75K}.
However, the temperature profile serves in our study only as a basis for determining the initial 
gas pressure of CSM (where $\rho$ may be either 
$\rho_\text{sw}$ or $\rho_\text{csd}$), 
\begin{align}\label{soundisosquared}
P=\rho a^2,\quad\text{ where }\quad a^2=kT/(\mu m_\text{u}),
\end{align}
denoting $k$ the Boltzmann constant, $\mu$ the mean molecular weight,
and $m_\text{u}$ the atomic mass unit.
After doing various simulations using different indices $0\le p\le 0.4$ with a minor 
impact on the effect of the counterpressure,
we use in the simulations performed in the study the power-law index $p=0$ (cf. also Sect.~\ref{numero}). For the 
same reason in the calculations we neglect the disk velocities $V_R$ and $\Omega$ 
(which are negligible comparing to SN-CSM interaction dynamics) as well as the effects 
of gravitational force and viscosity.

\begin{figure} [t]
\centering\resizebox{0.8\hsize}{!}{\includegraphics{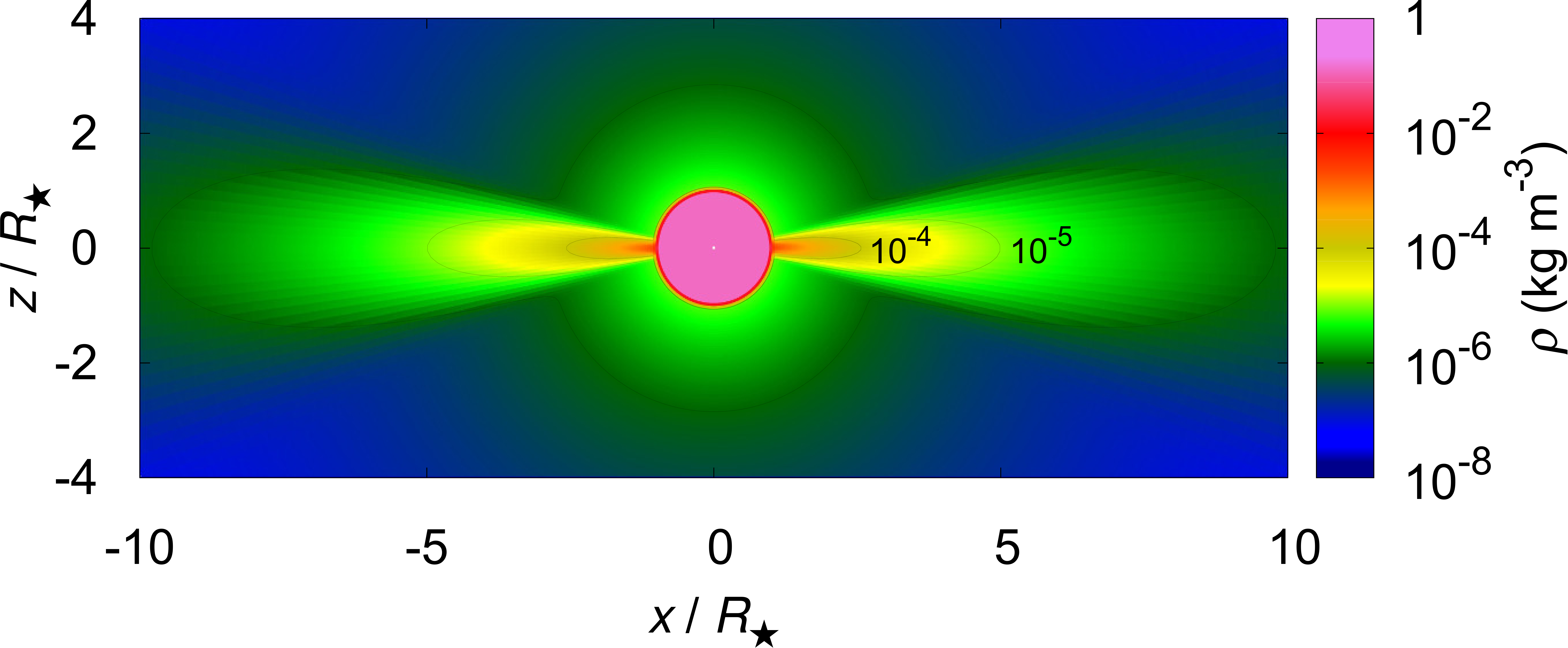}} 
\caption{Color map of the initial density of the SN progenitor
with $M_\star=45\,M_\odot$, $R_\star=80\,R_\odot$ (with mean density
$\langle\rho\rangle_\star\approx 0.125\,\text{kg m}^{-3}$) and CSM with assumed 
$\dot{M}_\text{sw}=10^{-1}\,M_\odot\,\text{yr}^{-1}$ 
and $\dot{M}_\text{csd}=10^{-4}\,M_\odot\,\text{yr}^{-1}$
(corresponding to Model A in Sect.~\ref{numero}).
The base density of the spherically symmetric part of CSM 
$\rho_{0,\text{sw}}\approx 7\times 10^{-6}\,\text{kg}\,\text{m}^{-3}$, 
while the equatorial disk base density
$\rho_{0,\text{csd}}\approx 4\times 10^{-4}\,\text{kg}\,\text{m}^{-3}$
(see also Tab.~\ref{table1} in Sect.~\ref{numero}).}
\label{fig1}
\end{figure}

For the analytical approximation of the disk midplane density profile
$\rho_\text{eq}(R)$ we use the equation which relates $\rho_\text{eq}$ to
the vertically integrated density $\Sigma(R)$ \citep[e.g.,][]{2011A&A...527A..84K},
\begin{align}\label{sigmanula}
\Sigma=\sqrt{2\pi}\,\rho_\text{eq}H.
\end{align}
Following the analytical approach of, for example, \citet{2001PASJ...53..119O},
$\Sigma(R)\sim R^{-2}$ and by denoting $H$ the disk vertical scale height, $H=a/\Omega$, 
in the Keplerian disk, we (analytically) obtain 
\begin{align}\label{radvertigo}
\rho_{\text{eq}}(R)\approx\rho_{0,\text{csd}}(R_{\star}/R)^{3.5},
\end{align}
where $\rho_{0,\text{csd}}$ is the disk base density 
and $H\sim R^{3/2}$ in the
disk Keplerian region \citep{2014A&A...569A..23K}.
Numerical models usually find the power $d$ of disk midplane radial density slope 
$\rho_{\text{eq}}\sim R^{-d}$ between $3$ - $4$, and
the disk base density $\rho_{0,\text{csd}}$ is, for example, in case of classical Be stars, estimated 
between $10^{-10}\,\text{kg}\,\text{m}^{-3}
$ to almost $10^{-6}\,\text{kg}\,\text{m}^{-3}$
\citep[see, e.g.,][]{2013A&A...553A..25G}. 
In case of much denser disks around, for instance, sgB[e]s, we assume disk base densities
up to the values of on the order of $10^{-4}\,\text{kg}\,\text{m}^{-3}$ 
\citep[cf.][]{2018A&A...613A..75K}.
The vertical profile of the thin circumstellar disk for $z\ll R$, 
which is the case for disks up to the distance of several 
stellar radii from the central object, takes the Gaussian form 
\citep[e.g.,][]{1981ARA&A..19..137P,2001PASJ...53..119O};
this reflects the disk vertical hydrostatic balance (see also Fig.~\ref{fig1}),
\begin{align}\label{thinvertigo}
\rho_\text{csd}(R,z)=\rho_\text{eq}(R)\,\text{e}^{\textstyle{-\frac{GM_{\star}}{a^2R^3}\frac{z^2}{2}}}. 
\end{align}

The initial structure of the disk is in this study basically parameterized by the disk mass-loss rate
$\dot{M}$. The (density independent) value of the disk base radial velocity
$V_R(R_\star)$ is pre-calculated from a 1D disk
model \citep[see][for more details]{2014A&A...569A..23K,kurfurst2015thesis}. The
disk base surface density $\Sigma_{0,\text{csd}}$ is fully determined from $\dot{M}$ and
$V_R(R_\star)$ using the equation of mass conservation Eq.~\eqref{mascon}.
The base surface density $\Sigma_{0,\text{csd}}$ is subsequently converted
to the base volume density $\rho_{0,\text{csd}}$ using the Gaussian vertical
hydrostatic equilibrium solution (see Eqs.~\eqref{sigmanula} and \eqref{thinvertigo}).
The complete initial disk profile in pressure and density is obtained from 
Eqs.~\eqref{temperature}, \eqref{soundisosquared}, \eqref{radvertigo}, and \eqref{thinvertigo}.

\section{Hydrodynamics and similarity solution of expanding supernova envelope}\label{sedov}
It is conventional to describe the evolution of a SN remnant
in several consecutive phases \citep[e.g.,][]{1999ApJ...510..379M}.
In the early phase the ejecta expands more or less ballistically into the surrounding medium. 
Ahead of the ejecta a strong shock is driven into the ambient medium while the
high pressure behind the forward shock drives a reverse shock into the ejecta 
\citep{1982ApJ...258..790C,1989ApJ...341..867C,1999ApJS..120..299T}. 
This ``ballistic'' or free expansion phase ends when the total amount of material 
swept up by the forward shock has a mass comparable to that of the ejecta, 
or equivalently, the mean density of the ejecta drops to that of the surrounding medium. 

The basic analytical solution of the SN explosive event follows the idea of homologous ($u\propto r/t$) 
energy-conserving hydrodynamic expansion, using the Sedov-Taylor blast wave 
solution \citep{1959sdmm.book.....S}
and the Sakurai solution of a shock wave in a nonuniform medium \citep{Sakurai} that has different density 
profiles of an internal gas and surrounding medium 
\citep{1977MoIzN....Q....S,1967pswh.book.....Z}. 
Gravitational force is unimportant except for the very interior layers, and
the radiative losses are almost negligible in the early phase of expansion when 
comparing to a kinetic energy of the gas. We thus may regard the process as adiabatic.
The scaling parameters are the energy of explosion $E_\text{SN}$, mass $M_\text{ej}$ 
of the expanding ejecta, and initial stellar radius $R_\star$. 

The main aim of our study 
is the modeling of interactions between the spherically expanding SN remnant
and the surrounding CSM, 
which may be spherically symmetric (stellar winds and/or pre-explosion gaseous shells)
or asymmetric (circumstellar disks, disk-like density enhancements, and bipolar lobes). Therefore 
we neglect a pre-explosion SN interior density distribution, 
which is completely transformed during the explosion, and on the other hand 
is of little influence on the structure of the studied interaction. 
We also do not introduce the complete analytical formalism in this section,
and we only briefly review the main relations in the way in which we use these for the 
comparative semi-analytical solution 
given in Sect.~\ref{semianal}. For the complete description see the fundamental solutions of, for example, 
\citet{1982ApJ...258..790C} and \citet{1985Ap&SS.112..225N}.

The basic hydrodynamic equations in an
axially symmetric slice of spherical coordinates in the plane $\phi=0$,
where the angle $\theta$ ranges therefore in full angle to include the plane $\phi=\pi,$ 
and where $x=R\sin\theta,\,z=R\cos\theta,\,\theta\in\langle 0,2\pi\rangle$ 
 (involving only the terms that are fundamental for nonviscous, non-self-gravitating,
and axially symmetric ($\partial/\partial\phi=0$)
gas expansion), i.e., the continuity equation, radial and polar components of momentum 
equation, and energy equation, take the form
\begin{gather}
\label{mombase01}\frac{\partial\rho}{\partial t}+\frac{1}{r}
\frac{\partial}{\partial r}\left(r\rho u_r\right)+
\frac{1}{r}\frac{\partial\left(\rho u_\theta\right)}{\partial\theta}= 0,\\[3pt]\label{mombase11}
\frac{\partial u_r}{\partial t}+u_r\frac{\partial u_r}{\partial r}+
\frac{u_\theta}{r}\frac{\partial u_r}{\partial\theta}-\frac{u_\theta^2}{r}+
\frac{1}{\rho}\frac{\partial P}{\partial r}-F_0= 0,\\[3pt]\label{mombase012}
\frac{\partial u_\theta}{\partial t}+u_r\frac{\partial u_\theta}{\partial r}+
\frac{u_\theta}{r}\frac{\partial u_\theta}{\partial\theta}
+\frac{u_r u_\theta}{r}+\frac{1}{\rho r}\frac{\partial P}{\partial\theta}
=0,\\[3pt]\label{mombase21}
\frac{\partial\left(P\rho^{-\gamma}\right)}{\partial t}+
u_r\frac{\left(P\rho^{-\gamma}\right)}{\partial r}
+\frac{u_\theta}{r}\frac{\left(P\rho^{-\gamma}\right)}{\partial\theta} = 0,
\end{gather}
where $\rho$ is the density, $u_r$ is the radial component of the flow velocity, 
$u_\theta$ is the polar component of the flow velocity, $P$ is the scalar pressure, 
$F_0$ is the body force per unit mass (typically gravitation),
and $\gamma$ is the ratio of specific heats.
Equation \eqref{mombase21} is obtained from the zero total derivative of the adiabatic transformation of a perfect gas
$P\rho^{-\gamma} = \text{const.}$
However, we further neglect the gravity term $F_0$
in the momentum equation \eqref{mombase11}, since gravitational 
force is of very little importance for the studied process.

In the reference frame that is co-moving with the forward shock
(for simplicity the planar propagation of the very thin shocked layer)
omitting viscosity and assuming constant adiabatic index
$\gamma$, which is astrophysically relevant \citep{1985Ap&SS.112..225N}, 
we can rewrite the adiabatic hydrodynamic equations \eqref{mombase01} - \eqref{mombase21} 
on both sides of the shock front into the simple conservative form \citep{1967pswh.book.....Z},
\begin{gather}\label{zelda}
\rho_1 u_1 = \rho_0u_0,\\[3pt]
\rho_1u_1^2+P_1 = \rho_0u_0^2+P_0,\\[3pt]
\frac{\gamma}{\gamma-1}\frac{P_1}{\rho_1}+\frac{u_1^2}{2} = 
\frac{\gamma}{\gamma-1}\frac{P_0}{\rho_0}+\frac{u_0^2}{2},
\end{gather}
where the upstream and downstream quantities are labeled with
subscripts 0 and 1, respectively. The same velocities in the laboratory frame are $D-u_0$ and $D-u_1$, where we denote 
$D$ the propagation speed of the forward shock.
To get the analytical solution,
in the early phase after explosion we approximate the process as a free
expansion with the internal velocity $u$ of the gas proportional to radius. Assuming the density profile 
$\rho_{\text{in}}$ in the
envelope as a (time-dependent) power law and the density of the surrounding (initially static)
medium $\rho_{\text{out}}$ as an alternate power law, we obtain 
\begin{align}\label{homolog1}
u=\frac{r}{t},\quad\rho_{\text{in}}(t)=Ar^{-n}t^{n-3},\quad\rho_{\text{out}}=Br^{-w}.
\end{align}
Assuming naturally the strong shock condition, $P_{\text{out}}\ll P_{\text{in}}$, 
the proportionality for the inner pressure, $P_{\text{in}}\sim \rho_{\text{in}}u^2$ 
\citep[in the shock co-moving frame, or alternatively $P_{\text{in}}\sim \rho_{\text{out}}D^2$ 
in the laboratory frame; see][]{1967pswh.book.....Z},
gives the inner and outer pressure in the form 
\begin{align}\label{homolog1aa}
 P_{\text{in}}(t)=\tilde{A}r^{2-n}t^{n-5},\quad P_{\text{out}}=\tilde{B}\rho_{\text{out}}^\gamma.
\end{align}
       
We describe the formalism of the analytical similarity solution based on Rankine-Hugoniot relations 
\citep[see the principles given in][]{1982ApJ...258..790C,1985Ap&SS.112..225N} 
in the Appendix \ref{basesimeqs}, 
while we describe the principles of semi-analytical model based 
on analytical solution in Appendix \ref{semianal}. 
We also describe in Appendix \ref{limanal} the analytical principles of the fundamental limitation of the adiabatic 
self-similar solution for only a certain time interval. 
The specific values of this limitation are given within the Sections corresponding to the particular models.

\section{Numerical method}\label{nummet}
For the modeling of the interaction with extremely sharp discontinuities characterized by 
extremely high pressure gradients we have 
developed and used our own version of single-step 
(unsplit, ATHENA-like \citep[e.g.,][]{2008ApJS..178..137S}) finite volume Eulerian hydrodynamic 
code based on Roe's method 
\citep[][see also the description of our geometrical modifications 
of the algorithm used for different astrophysical situations in 
\citealt{KKapplmath}]{1981JCoPh..43..357R,Toro,2018A&A...613A..75K}. 

For the time-dependent calculations we write the adiabatic hydrodynamic equations 
\eqref{mombase01} -- \eqref{mombase21} in two different manners. 
The primitive form generally is \citep[see, e.g.,][]{2008ApJS..178..137S}
\begin{align}\label{numappr1}
\frac{\partial{\vec{W}}}{\partial t}+\vec{\nabla}\cdot\vec{F}(\vec{W})=0, 
\end{align}
where the primitive variables are 
$\vec{W}=\rho,\,\vec{u},\,P$, and the fluxes 
$\vec{F}(\vec{W})=\rho\vec{u},\,\rho\vec{u}\,|\,\vec{u}+P\,\mathbf{1},\,\rho\vec{u}H$,
where $|$ denotes the dyadic product of two vectors,
$H=(E+P)/\rho$ is the enthalpy (+ specific kinetic energy),
$\mathbf{1}$ is the unit matrix, and $E=\rho\epsilon+\rho u^2/2$ is total energy 
(while $\epsilon$ is specific internal energy and $u^2=u_r^2+u_\theta^2$).
The primitive (as well as conservative) form of continuity equation is given by Eq.~\eqref{mombase01} ,
while Eqs.~\eqref{mombase11} and \eqref{mombase012} represent the primitive form 
of radial and polar components of momentum equation. We transform Eq.~\eqref{mombase21} into explicit 
primitive form of energy equation by inserting Eq.~\eqref{mombase01},
\begin{align}
\label{mombase211}
&\frac{\partial P}{\partial t}+u_r\frac{\partial P}{\partial r}+
\frac{u_\theta}{r}\frac{\partial P}{\partial\theta} 
+ \gamma P\left(\frac{\partial u_r}{\partial r}+\frac{u_r}{r}+
\frac{1}{r}\frac{\partial u_\theta}{\partial\theta}\right)= 0,
\end{align}
where $\gamma=4/3$ for the radiation dominated gas.
Equations \eqref{mombase01} - \eqref{mombase012} together with 
Eq.~\eqref{mombase211} are used in the primitive numerical scheme (Eq.~\eqref{numappr1}).

The analogous conservative form is 
\citep[see, e.g.,][]{1986ASIC..188..187N,1989nyjw.book.....H,1992ApJS...80..753S,1995A&A...299..523F,
LeVeque1998Comput}
\begin{align}\label{numappr2}
\frac{\partial{\vec{U}}}{\partial t}+\vec{\nabla}\cdot\vec{F}(\vec{U})=0,
\end{align}
where the conservative variables are 
$\vec{U}=\rho,\,\vec{M},\,E$ and $\vec{F}(\vec{U})=
\rho\vec{u},\,\vec{M}\,|\,\vec{u}+P\,\mathbf{1},\,\vec{M}H$ (where $\vec{M}=\rho\vec{u}$ 
is the momentum density) for the mass, momentum, and energy equations, respectively. 
Abbreviating $\gamma^\prime=\gamma/(\gamma-1)$, the explicit adiabatic form of the 
enthalpy $H$ used in the flux terms is $H=\gamma^\prime P/\rho+u^2/2$.
The explicit conservative form of radial (omitting the gravitational force; see Sect.~\ref{sedov}) 
and polar components of momentum equation and of the energy equation, respectively, is
\begin{align}\label{conservativemomr}
&\frac{\partial M_r}{\partial t}+\frac{1}{r}\frac{\partial}{\partial r}\left(rM_ru_r\right)+
\frac{1}{r}\frac{\partial(M_\theta u_r)}{\partial\theta}
-\rho\frac{u_\theta^2}{r}+
\frac{\partial P}{\partial r}
= 0,\\[3pt]\label{conservativemomtheta}
&\frac{\partial M_\theta}{\partial t}+\frac{1}{r}\frac{\partial}{\partial r}\left(rM_ru_\theta\right)+
\frac{1}{r}\frac{\partial(M_\theta u_\theta)}{\partial\theta}
+\rho\frac{u_r u_\theta}{r}+\frac{1}{r}\frac{\partial P}{\partial\theta}=0,\\[3pt]
\label{conservativeenerg}
&\frac{\partial E}{\partial t}+\frac{1}{r}\frac{\partial}{\partial r}\left(r M_r H\right)+
\frac{1}{r}\frac{\partial}{\partial\theta}\left(M_\theta H\right)= 0.
\end{align}
Equation \eqref{mombase01} together with Eqs.~\eqref{conservativemomr} - \eqref{conservativeenerg} 
are used in the conservative numerical scheme (Eq.~\eqref{numappr2}).

The polar form of Roe matrices $\vec{A}_{\vec W}=\partial\vec{F}(\vec{W})/\partial{\vec W}$ 
and $\vec{A}_{\vec U}=\partial\vec{F}(\vec{U})/\partial{\vec U}$, 
resulting from the linearization of equations \eqref{numappr1} and \eqref{numappr2},
is in this case identical and is supplemented by formalism introduced in
\citet{2008ApJS..178..137S} and \citet{2010ApJS..188..290S}.
We used the (fully capable) second-order piecewise linear reconstruction algorithm, 
without any necessity to alternate the Roe's method using the HLLE 
(Harten, Lax, van Leer, and Einfeldt; cf.~\citet{1991JCoPh..92..273E,2008ApJS..178..137S}) 
solver or similar positive-definite solvers in case of unexpected density or pressure drop to negative values in the 
intermediate states of computation \citep[see][for the details]{2008ApJS..178..137S}, even in extremely thin 
zones of very strong shocks.

To derive the initial conditions for hydrodynamic quantities at the explosion
time $t_0$ we assumed that the interior shock wave basically rearranges the original 
density structure of the progenitor. 
The density gradient between the inner and outer interstellar region is reduced by the rapid expansion of
the inner regions \citep[][see also, e.g., Fig.~1 in \citealt{1986ninp.proc..363A} 
showing the evolution of the density structure from core collapse to mantle ejection]
{1996snai.book.....A}. 
We therefore neglect the details in a SN interior density structure 
that can barely affect the main features of examined effects in a complex 
interaction between SNe and CSM. We thus assume for simplicity 
that the stellar density is equal to the ``averaged''
homogeneous initial density $\langle\rho\rangle_{\!\star}=3M_\star/(4\pi R_\star^3)$ of an exploding star 
(according to the stellar parameters introduced in Sects.~\ref{sedov}, \ref{numero}, and \ref{semianal}) 
in the range between $0.1-1\,\text{kg}\,\text{m}^{-3}$ 
($\langle\rho\rangle_{\!\star}\approx 0.125\,\text{kg}\,\text{m}^{-3}$; cf.~Fig.~\ref{fig1}). 

The injected SN explosion energy is $E_\text{SN}=10^{44}\,\text{J}$. 
A SN explosion drives a radiation dominated shock, i.e., the internal energy inside 
the sphere of shock wave that propagates through the star
is dominated by radiation \citep{2010ApJ...725..904N}. For that reason
we assume the averaged homogeneous initial internal pressure of a photon gas;
we consider the whole stellar volume as a ``thermal bomb'' and neglect the details of 
the processes connected with the shock wave propagation through the star which, 
in fact, have a secondary effect on the studied interaction 
of SN with the aspherical CSM. The averaged homogeneous initial internal pressure of a photon of gas takes the form, 
$\langle P\rangle_{\!\star}=E_\text{SN}/(3V_{\!\star})=E_\text{SN}/(4\pi R_\star^3)$, 
i.e., in the range between 
$10^{10}$ - $10^{11}\,\text{Pa}$ ($\langle P\rangle_{\!\star}\approx 4.63\times 10^{10}\,\text{Pa}$).

We insert the initial state for the CSM in the following explicit form: the density of a 
spherically symmetric CSM is expressed simply by Eq.~\eqref{spheresym} while its
outflow velocity
$\varv_\text{sw}=100\,\text{km}\,\text{s}^{-1}$ \citep[see Sect.~\ref{windbaseq}
and also][]{2014MNRAS.439.2917M} 
is used only for initial density estimate and is neglected for the calculation itself. 
We assume an isothermal initial temperature structure of the 
CSM (of both the components) within the computational domain, given as 
$T_\text{sw}=T_\text{csd}=0.7T_\text{eff}$ 
\citep[cf.][]{2008ApJ...684.1374C,2018A&A...613A..75K}. 
This also indicates the initial counterpressure of the CSM as $P=\rho a^2$, where 
the square sound speed $a^2$ is given by Eq.~\eqref{soundisosquared}.
The initial circumstellar equatorial disk density structure 
(Eqs.~\eqref{radvertigo} and \eqref{thinvertigo})
is in the polar coordinates specified in the following explicit form as
\begin{align}\label{diskvertdensity}
 \rho_\text{csd}(r,\theta)=\rho_{0,\text{csd}}\left(\frac{R_\star}{r\,|\text{sin}\,\theta|}\right)^{3.5}
 \exp\left(-\frac{GM_\star}{2a^2r}\frac{\cos^2\theta}{|\text{sin}^3\theta|}\right).
\end{align}
We neglect the velocity field of the circumstellar equatorial disk (see Sect.~\ref{circbaseq}), since we regard its effects 
during the studied process as negligible. 
The total CSM density profile is given as $\rho_\text{sw}\text{ (Eq.~\eqref{spheresym}) }+
\rho_\text{csd}$ (Eq.~\eqref{diskvertdensity}).

Within our considerations
we may set $R_0=R_\star$, while $t_0$ is certainly less than one day, 
for example, for SN 1987A the time $t_0$ is estimated to be 53 minutes 
after core collapse \citep{1996snai.book.....A}. 
We set the free boundary conditions at the inner (originally interstellar region) as well as at the outer boundary 
of the computational domain during the expansion phase. 
The values for the grid boundary and ghost zones are extrapolated from mesh interior values as a 
$0\text{th}$ order extrapolation.

We performed all the calculations on a center-symmetric area domain (2D axially symmetric slice 
of the spherically symmetric 3D domain) with a radius of $20\,R_\star$, and with a numerical polar
grid uniformly scaled in both directions $r$ and $\theta$. 
The star is located in the center of the computational domain.
We also tested the Cartesian uniform grid, 
interlaced by an equatorial plane of the spherical coordinate system, for comparison with the calculated 
model; however, we achieved smoother profiles and the optimal computational cost
using the polar grid.
The number of spatial grid cells was $400$ in radial and $300$ in azimuthal directions
for the models performed up to radius $20\,R_\star$ and in $2\pi$ polar domain.
The complete physical time of the simulations corresponds 
in this case to approximately 165 hours. We also performed the calculation on 2D grid containing one polar quadrant ($\pi/2$ polar domain) 
with 2400 grid cells in radial and 480 grid cells in azimuthal 
direction with the complete physical time of the simulations corresponding 
to approximately 180 hours. The selected
grid aspect ratio has a stabilizing effect because the cells that are 
stretched along the tangent to the shock have a damping effect 
on the development of the carbuncle numerical instability \citep[][]{2001JCoPh.166..271P}.
All the models are described in Sect.~\ref{numero} 
\citep[see also][]{2017ASPC..508...17K,KKapplmath,2018A&A...613A..75K}.
\begin{figure} [t!]
\begin{center}
\centering\resizebox{0.725\hsize}{!}{\includegraphics{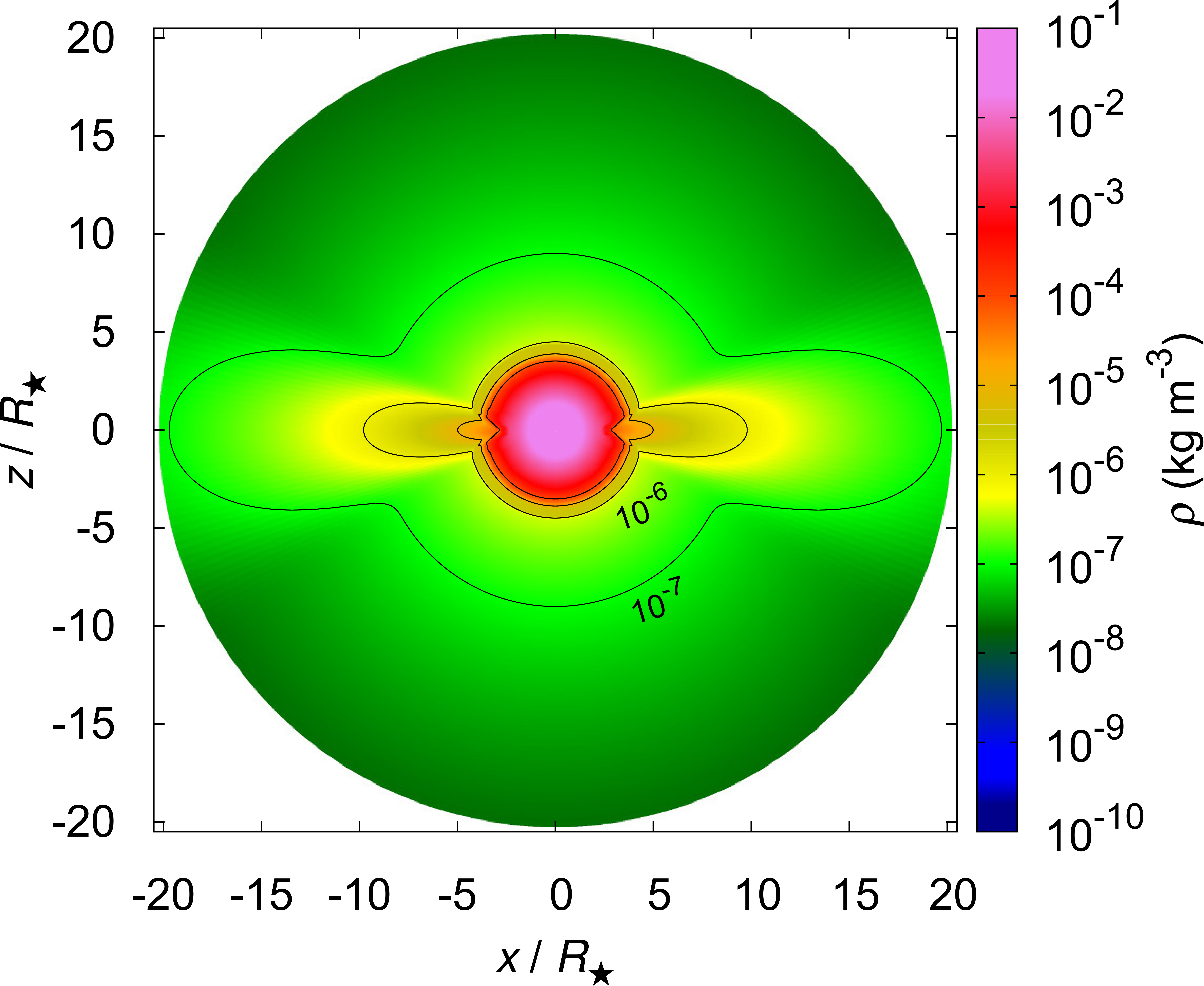}} 
\caption{\ Model A:
Color map of the density structure of the interaction of SN (with progenitor and CSM
parameters introduced in Sect.~\ref{sedov}, Tab.~\ref{table1}, and in Fig.~\ref{fig1}) 
with asymmetric CSM that forms a dense equatorial disk, at time 
$t \approx 16\,\text{h}$ since shock emergence. 
Contours denote the densities $\rho = 10^{-7},\,10^{-6},\,10^{-5},\text{ and }
10^{-4}\,\text{kg}\,\text{m}^{-3}$.}
\label{colorfig1}
\end{center}
\end{figure}
\begin{table}[t!]
\caption{Parameters of the three models A, B, and C.}
\label{table1}
\centering
\renewcommand{\arraystretch}{1.2}
\begin{tabular}{p{1.5cm} p{1.5cm} p{1.5cm} p{1.5cm}}
\multicolumn{4}{l}{Model A - high density of the surrounding CSM}\\
\hline
$\dot M_\text{sw}$ & $\dot M_\text{csd}$ & ~~$\rho_{0,\text{sw}}$ & ~$\rho_{0,\text{csd}}$\\\hline 
$10^{-1}$ & $10^{-4}$ & $7\times 10^{-6}$ & $4\times 10^{-4}$\\
\hline\\
\multicolumn{4}{l}{Model B - intermediate density of the surrounding CSM}\\
\hline
$\dot M_\text{sw}$  & $\dot M_\text{csd}$ & ~~$\rho_{0,\text{sw}}$ & ~$\rho_{0,\text{csd}}$\\\hline
$10^{-2}$ & $10^{-4}$ & $7\times 10^{-7}$ & $4\times 10^{-4}$\\
\hline\\
\multicolumn{4}{l}{Model C - low density of the surrounding CSM}\\
\hline
$\dot M_\text{sw}$  & $\dot M_\text{csd}$ & ~~$\rho_{0,\text{sw}}$ & ~$\rho_{0,\text{csd}}$\\\hline
$10^{-4}$ & $10^{-6}$ & $7\times 10^{-9}$ & $4\times 10^{-6}$\\
\hline
\end{tabular}
\tablefoot{Mass-loss rates $\dot M_\text{sw}$ and $\dot M_\text{csd}$
are in units of $M_\odot\,\text{yr}^{-1}$ and the (approximate) base densities
$\rho_{0,\text{sw}}$ and $\rho_{0,\text{csd}}$ are in SI units
($\text{kg}\,\text{m}^{-3}$).}
\end{table}

\section{Numerical models}\label{numero}
{We selected a BSG star with $M_\star=45\,M_\odot$, $R_\star=80\,R_\odot$, $T_\text{eff}=25\,000\,\text{K}$
as the progenitor for all studied models; 
all of the parameters of this star are described in Sect.~\ref{nummet} (see also Fig.~\ref{fig1}). 
The assumed properties of CSM of studied models (denoted as
Model A, B, and C), i.e., the adopted mass-loss rates $\dot M$ and the
approximate base densities $\rho_{0,\text{sw}}$ and $\rho_{0,\text{csd}}$
derived from the mass conservation equation (see Sects.~\ref{windbaseq} and
\ref{circbaseq}), are given in Table~\ref{table1}.
In all models we assume
$\varv_\text{sw}=100\,\text{km}\,\text{s}^{-1}$ 
\citep[][see also Sect.~\ref{nummet}]{2014MNRAS.439.2917M}, 
which is an order of magnitude 
latitudinally averaged estimate considering the effects of possible high rotation 
\citep{2000A&A...361..159M,2008A&A...487..697K,2010A&A...517A..30K} and high radiation pressure.}
\subsection{Model A: High density of the surrounding media}\label{numerodense}
\begin{figure} [t!]
\begin{center}
\centering\resizebox{0.725\hsize}{!}{\includegraphics{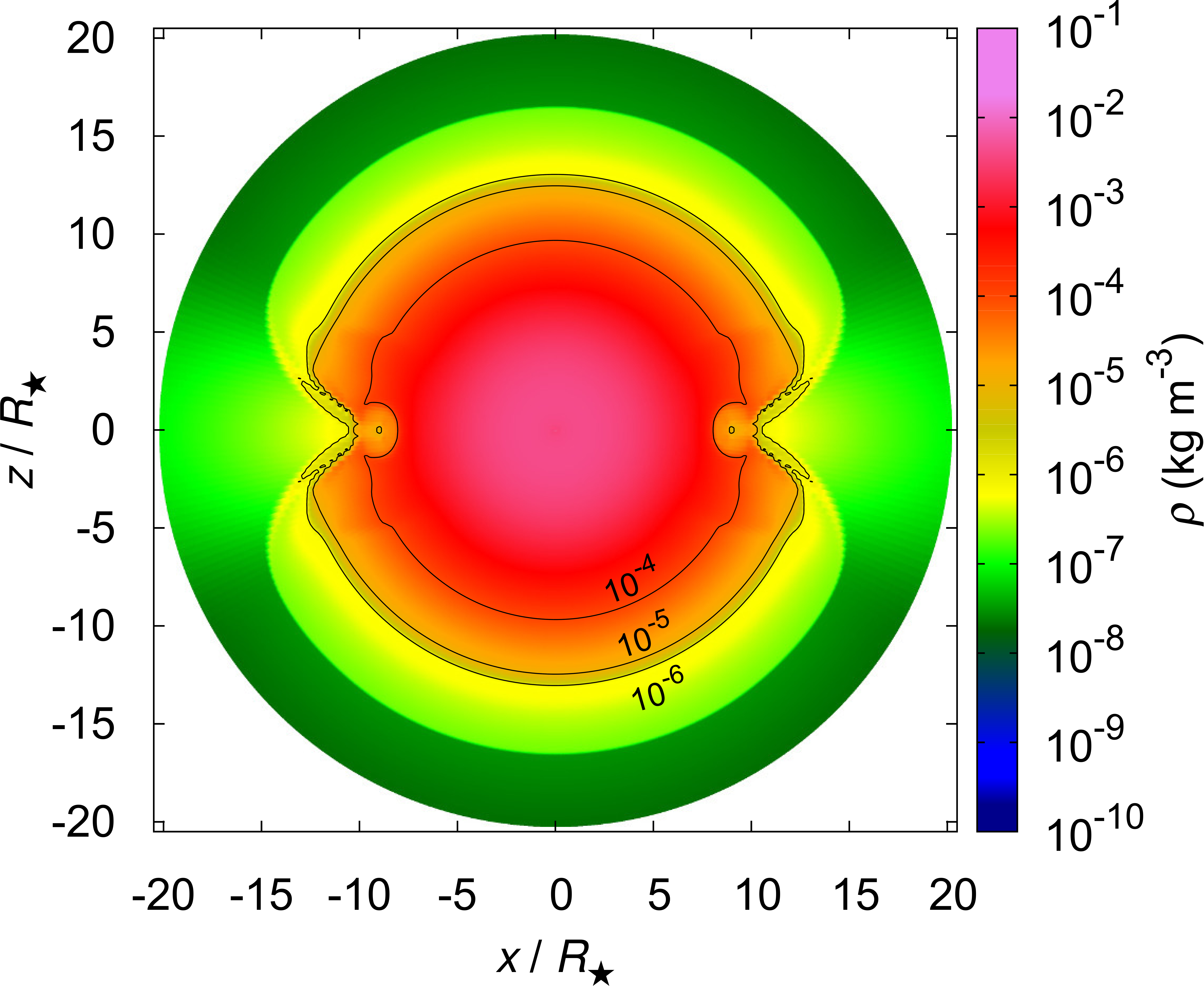}} 
\centering\resizebox{0.725\hsize}{!}{\includegraphics{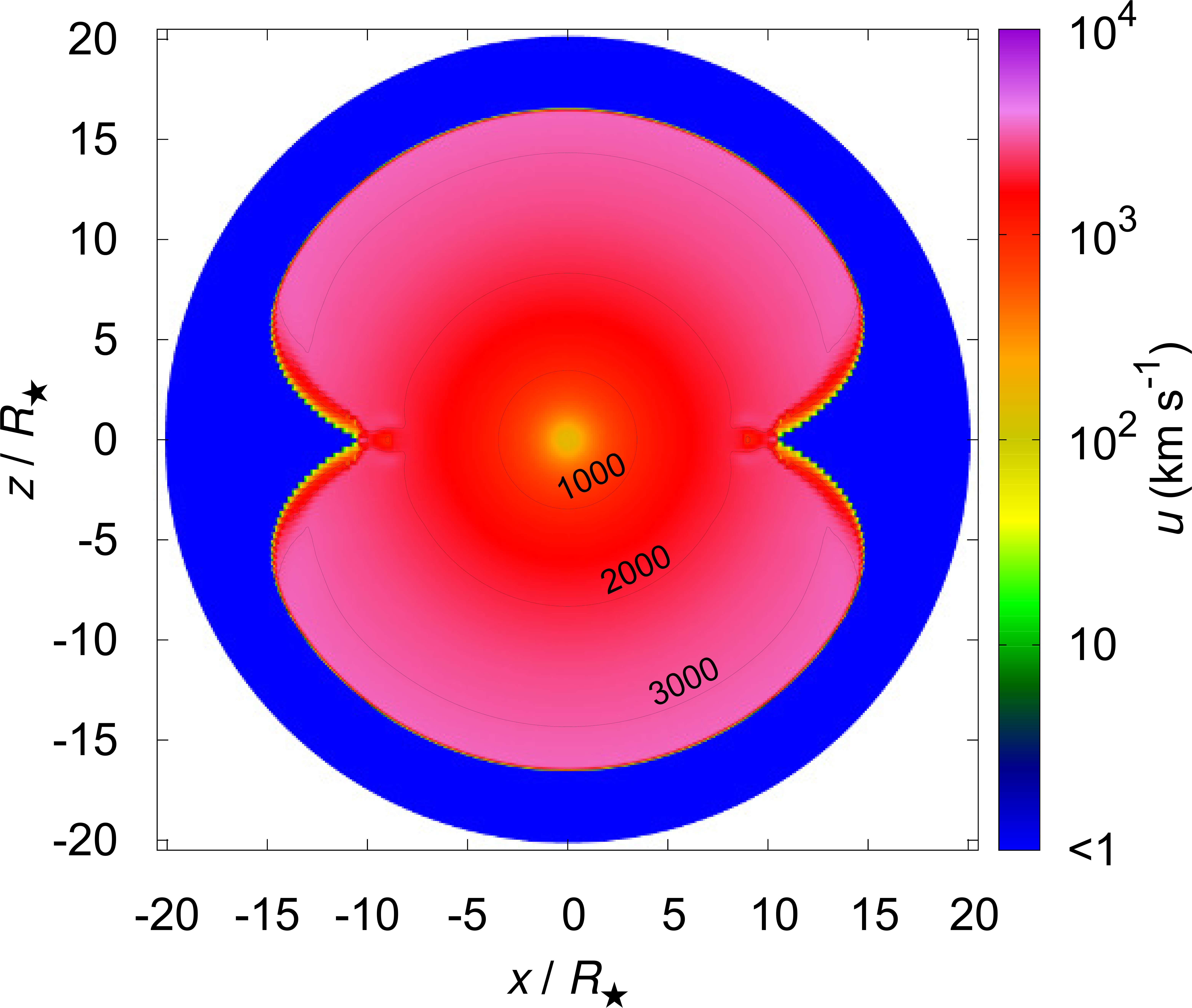}}
\centering\resizebox{0.725\hsize}{!}{\includegraphics{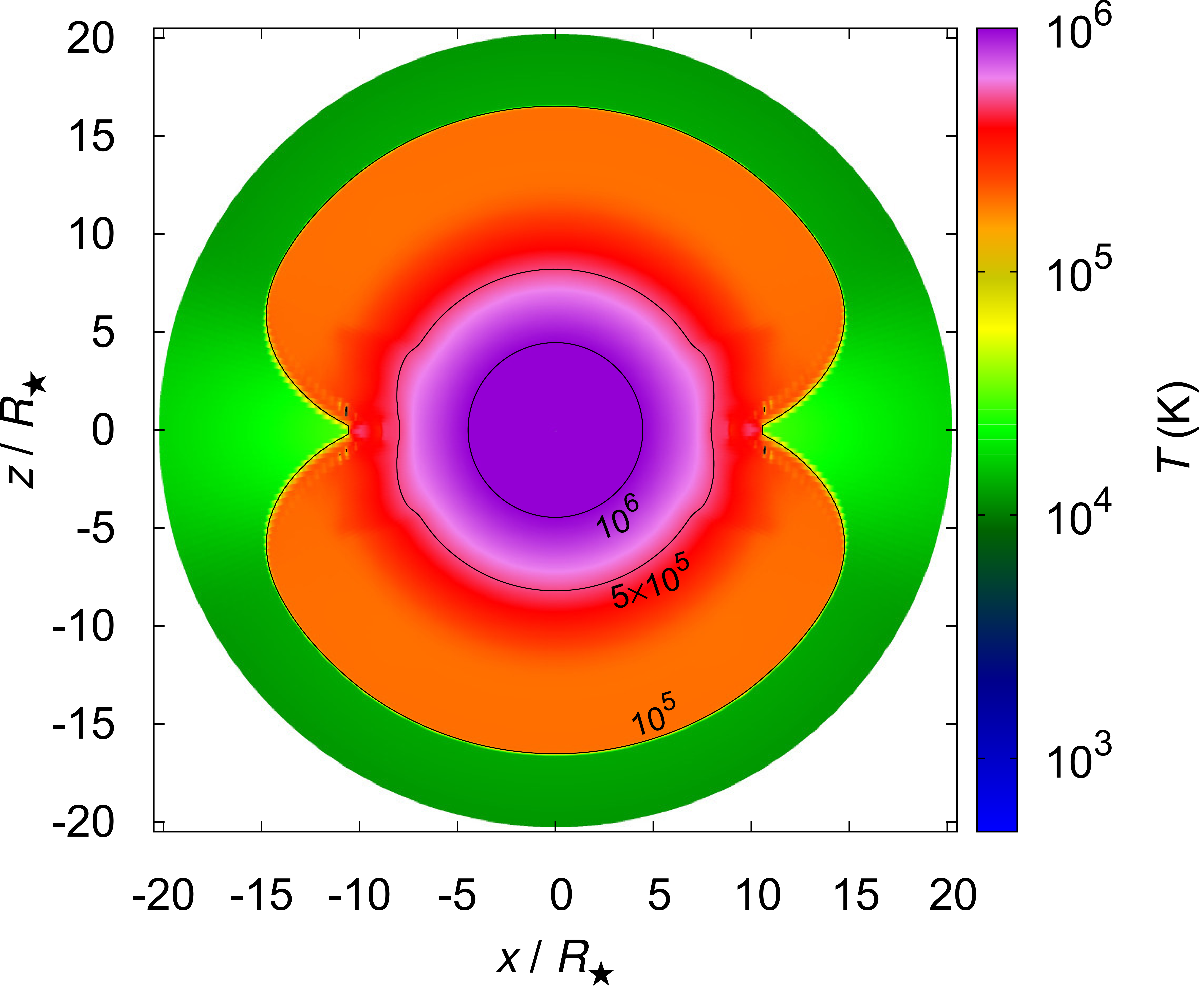}} 
\caption{Model A: Interaction of SN with asymmetric CSM at time 
$t \approx 66\,\text{h}$ since shock emergence.
\textit{Upper panel}: Color map of the density structure.
Contours denote the densities $\rho = 10^{-6},\,10^{-5},\text{ and }10^{-4}\,\text{kg}\,\text{m}^{-3}$.
\textit{Middle panel}: Color map of the velocity structure.
Contours denote the radial (expansion) velocities
$u = 1000$, $2000$, and $3000\,\text{km}\,\text{s}^{-1}$.
\textit{Lower panel}: Color map of the temperature structure.
Contours denote the temperatures $T = 10^{5},\,5\times 10^{5},\text{ and }10^{6}\,\text{K}$.
Characteristic 1D sections of the model in the equatorial and polar plane are shown in Fig.~\ref{fig2_all}.}
\label{colorfig2} 
\end{center}
\end{figure}
\begin{figure} [t!]
\begin{center}
\centering\resizebox{0.725\hsize}{!}{\includegraphics{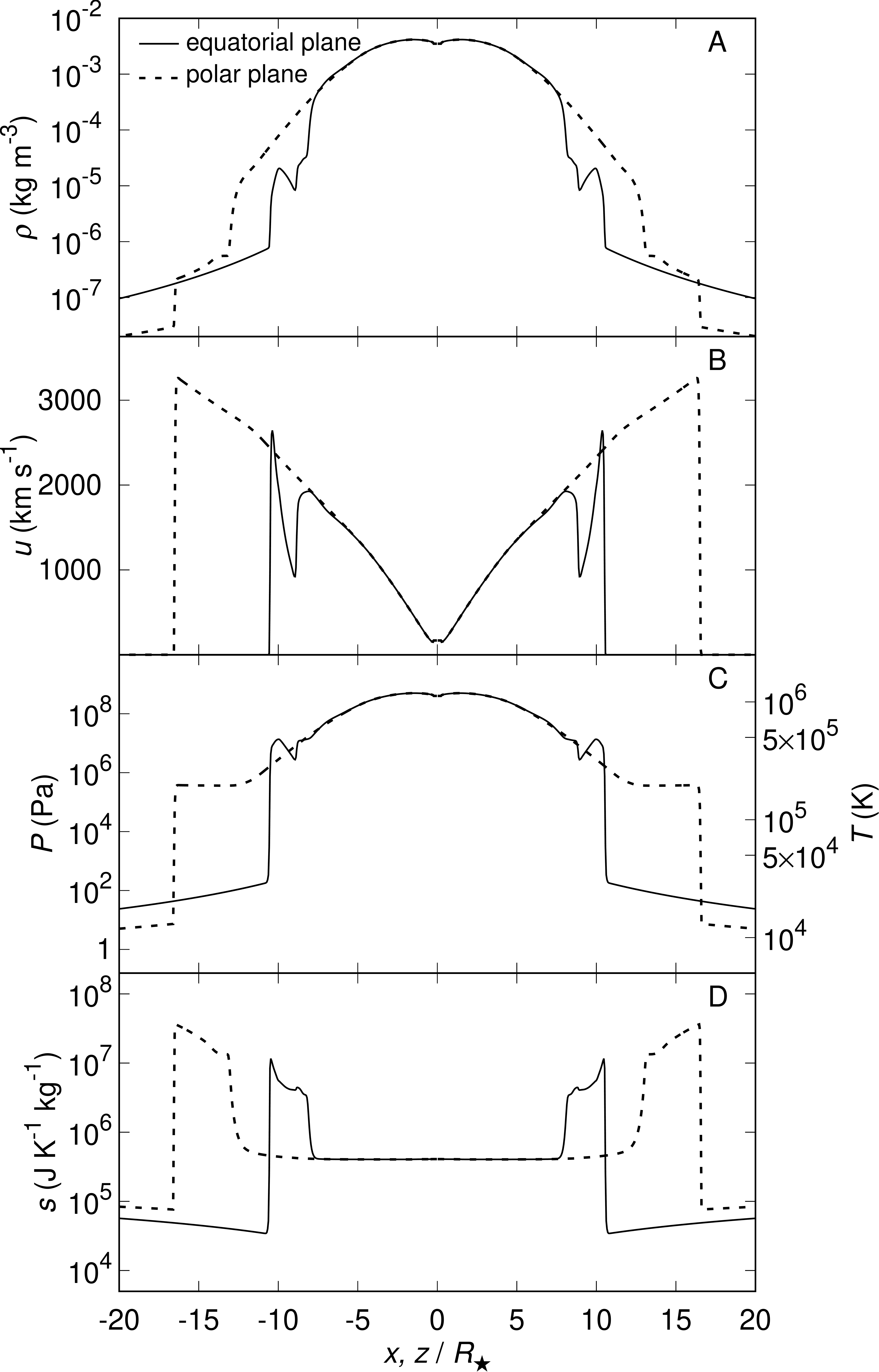}} 
\caption{Model A: One-dimensional slices of variations of hydrodynamical variables in the equatorial plane
($x$-coordinate, solid line) and in the polar direction
($z$-coordinate, dashed line), at the same time as in Fig.~\ref{colorfig2}.
\textit{Panel A}: Density $\rho$.
\textit{Panel B}: Expansion velocity $u$.
\textit{Panel C}: Pressure $P$ and temperature $T$.
\textit{Panel D}: Specific entropy $s$.}
\label{fig2_all}
\end{center}
\end{figure}
\begin{figure} [t!]
\begin{center}
 \centering\resizebox{0.725\hsize}{!}{\includegraphics{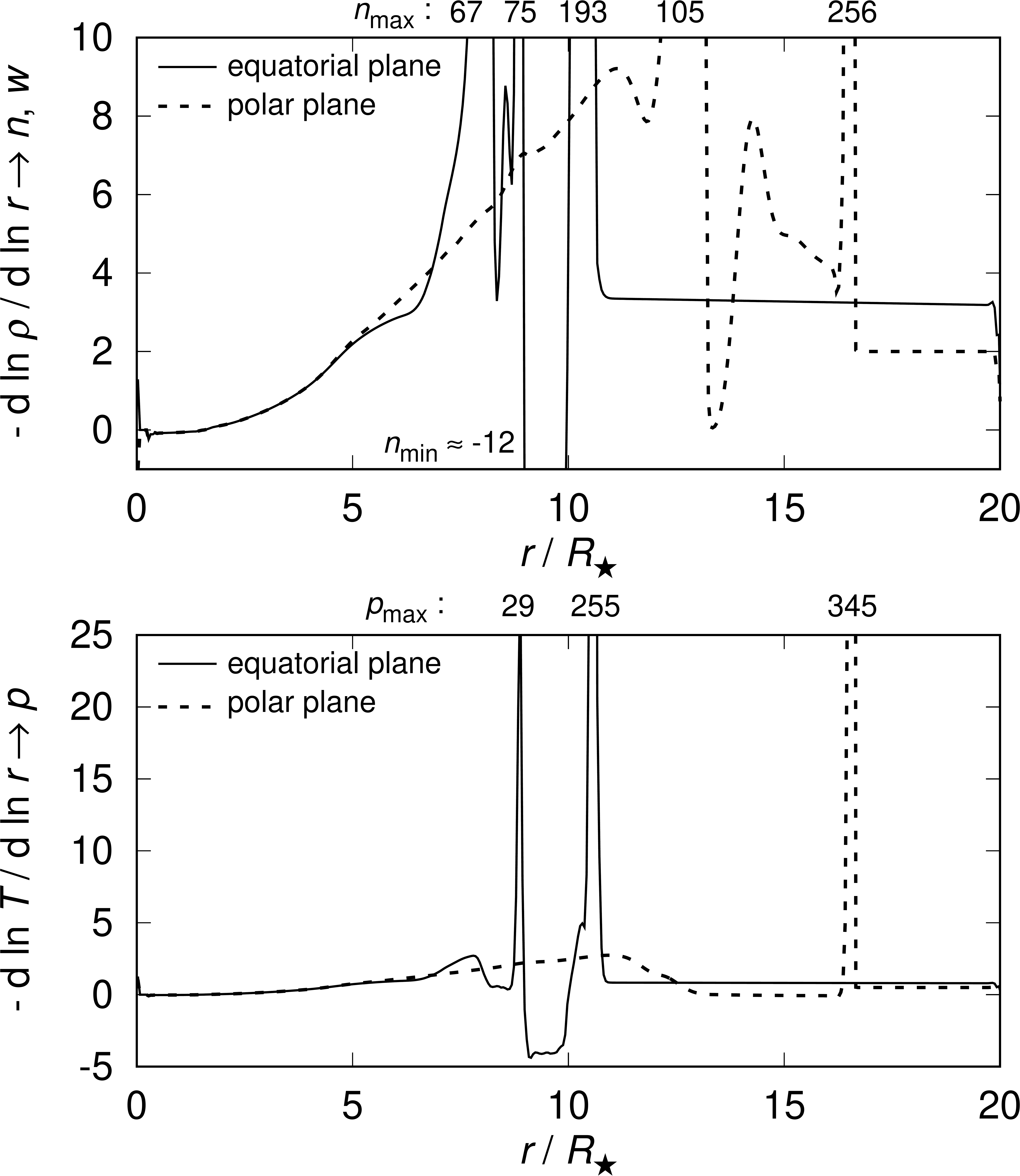}} 
 \caption{Model A: \textit{Upper panel}: Slope parameters $n$ and $w$ of the density 
(see Eq.~\eqref{homolog1}) in the equatorial (solid line) and polar (dashed line) direction
corresponding to the panel A in Fig.~\ref{fig2_all}, at the same time. 
The position of the outer shock wave is at 
 $r\approx 11\,R_\star$ in the equatorial direction, while it is at 
 $r\approx 16.5\,R_\star$ in the polar direction.
 There are preserved the initial density slope parameters of CSM, $w_\text{sw}=2$ 
 in the polar direction and $w_\text{csd}=3.5$ in the equatorial direction, 
 above these radii. 
 The inner envelope density slope parameter increases from $n=0$ to $n\approx 9$ in the polar direction. 
 The numbers $n_\text{max}$ and $n_\text{min}$ above and below the graph 
 denote the maximal and minimal slopes 
 within the peaks of the density gradient discontinuities.
 \textit{Lower panel}: Slope parameters $p$ of the temperature (cf.~Eq.~\eqref{temperature}) 
 in the equatorial (solid line) and polar (dashed line) direction
 corresponding to the temperature graph (panel C) in Fig.~\ref{fig2_all} at the same time. 
 The inner envelope (smooth) temperature slope parameter increases to $p\approx 2.7$ at 
 $r\approx 7.8\,R_\star$ in the equatorial direction while 
 it increases to approximately the same value at $r\approx 11\,R_\star$ in the polar direction. 
 The slope parameter $p$ outside 
 the outer shock is $p=0.875$ in the equatorial and $p=0.5$ 
 in the polar direction.
 The numbers $p_\text{max}$ above the graph denote the maximal temperature slopes within
 the peaks of the $T$ gradient discontinuities.}
 \label{fig33}
 \end{center}
\end{figure}
\begin{figure} [t!]
\begin{center}
 \centering\resizebox{0.96\hsize}{!}{\includegraphics{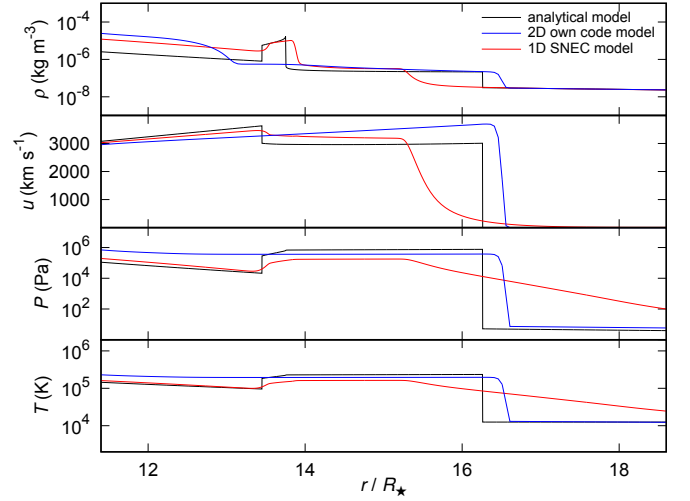}} 
 \caption{Model A:
 Results of semi-analytical solution of the density $\rho$, velocity $u$, pressure $P$, and temperature $T$ (black line), 
 with the slope parameters $n=7$ and $w_\text{sw}=2$, in the polar direction. 
 The method of calculation is described in Sect.~\ref{semianal}. 
 The time is the same as in Figs.~\ref{colorfig2} - \ref{fig33}. 
 The velocity at the outer shock wave region (at approx.~$16\,R_\star$) 
 corresponds to the polar expansion velocity in the panel B of 
 Fig.~\ref{fig2_all} while the other quantities and semi-analytical model input 
 variables are described in Sect.~\ref{semianal}.
 The blue line depicts the corresponding calculation from our 2D model.
 The corresponding 1D profile calculated using the SNEC-1.01 code 
 is depicted with a red line (see the description in Sect.~\ref{numerodense}).}
 \label{fig44}
 \end{center}
\end{figure}
\begin{figure} [t!]
\begin{center}
 \centering\resizebox{0.96\hsize}{!}{\includegraphics{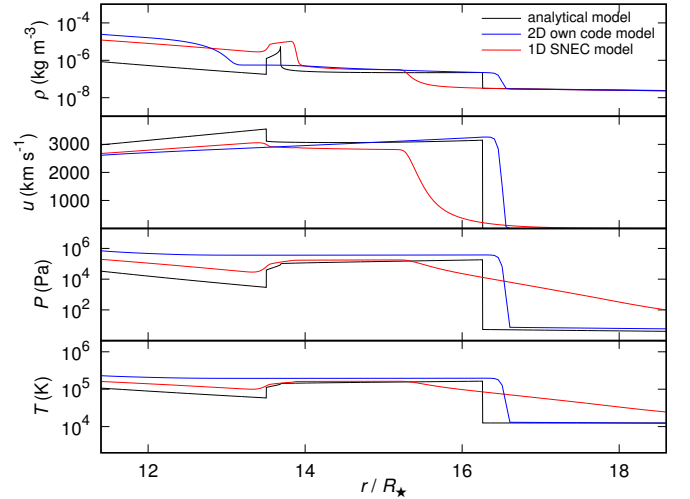}} 
 \caption{Model A: As in Fig.~\ref{fig44}, with the slope parameters 
 $n=9$ and $w_\text{sw}=2$, in the polar direction (the same method of calculation 
 as in Fig.~\ref{fig44}). The semi-analytical profiles are compared
to the same polar expansion velocity as in Figs.~\ref{fig2_all} and \ref{fig44}.}
 \label{fig55}
 \end{center}
\end{figure}

\begin{figure} [t!]
\begin{center}
\centering\resizebox{0.85\hsize}{!}{\includegraphics{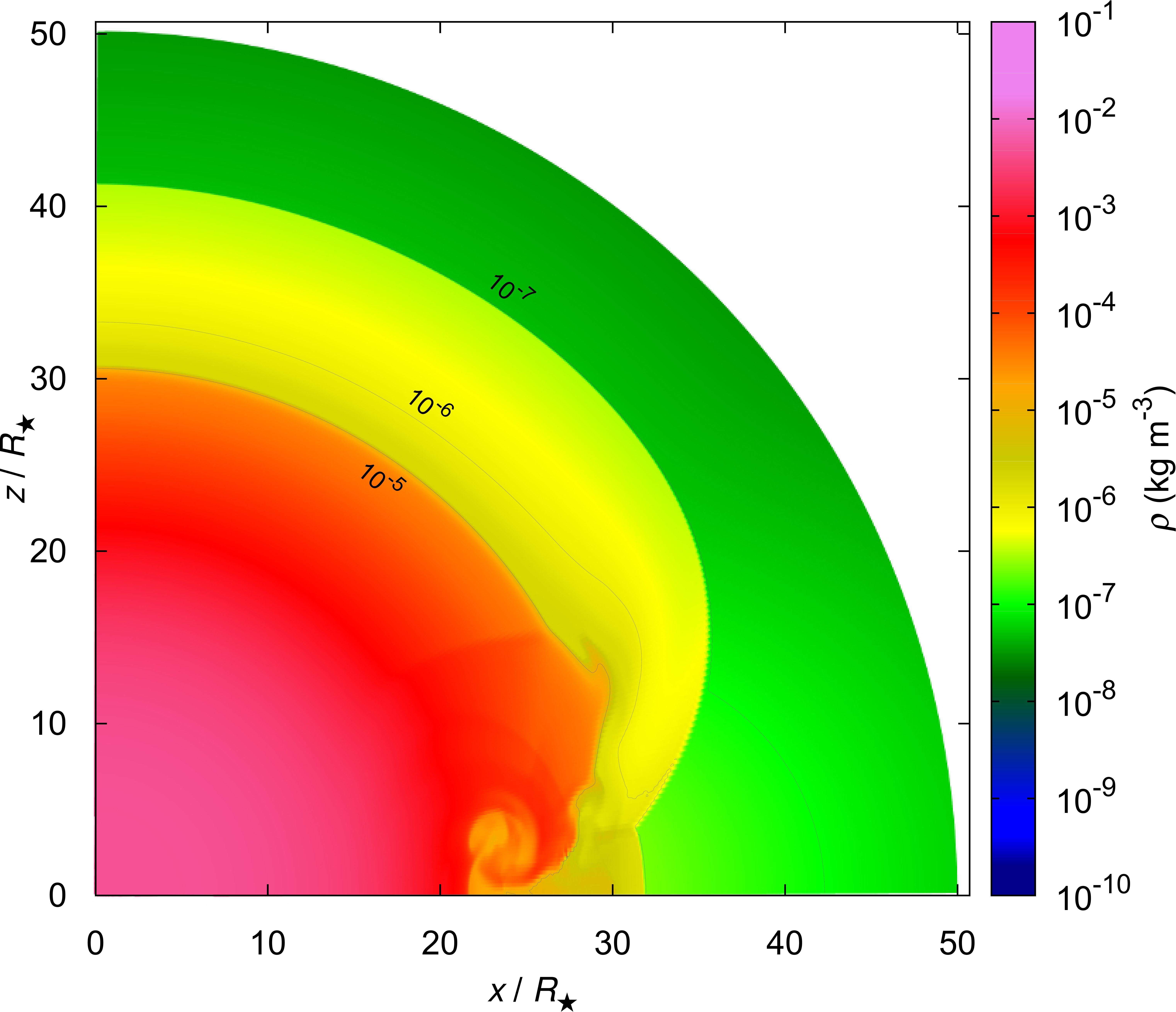}} 
\centering\resizebox{0.85\hsize}{!}{\includegraphics{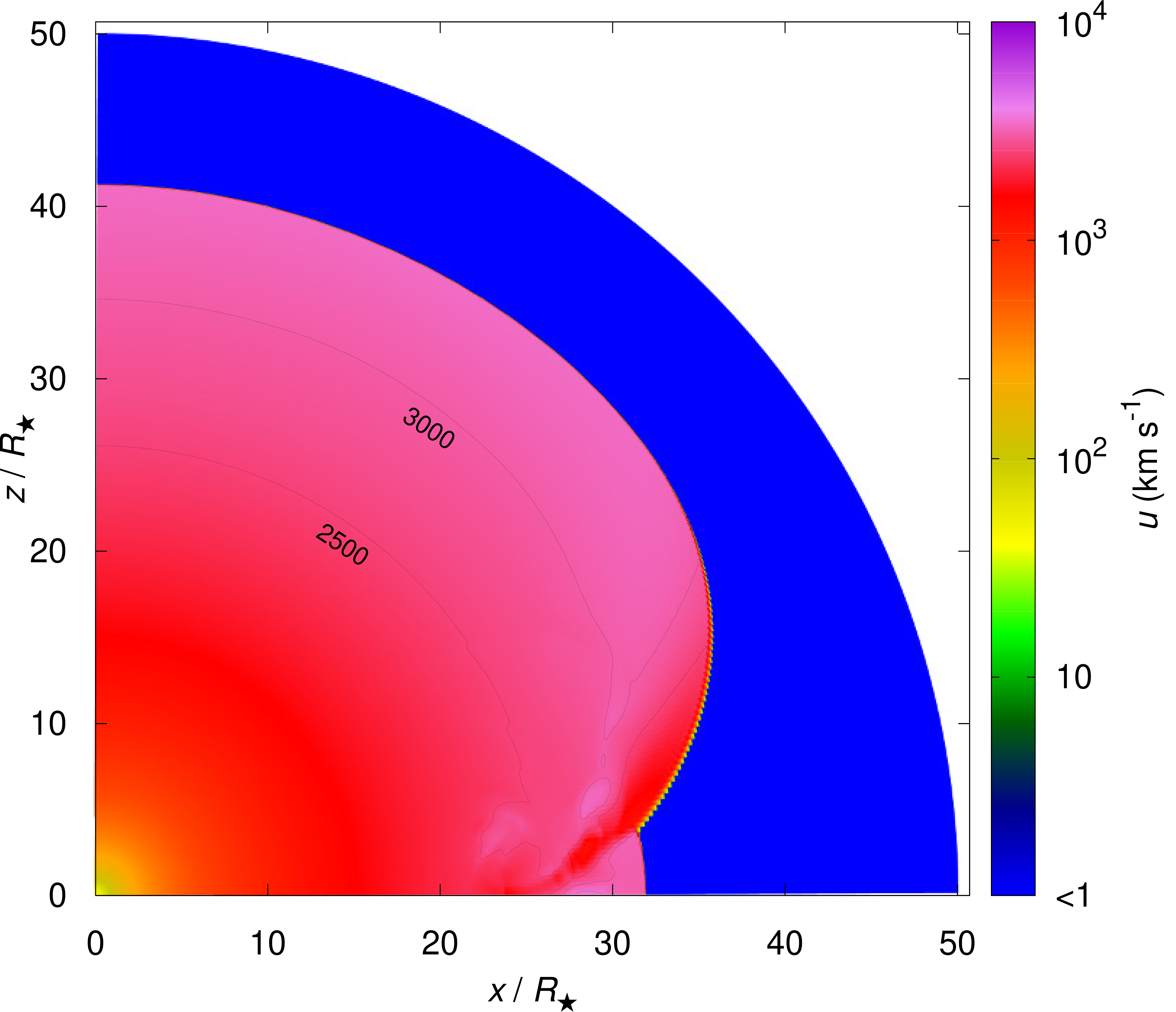}} 
\caption{Model A: Detailed color maps of the density (\textit{upper panel}) and the 
radial velocity (\textit{lower panel}) structure,
up to the distance $50\,R_\star$ at time $t \approx 165\,\text{h}$ since shock emergence.
Contours denote the densities $\rho = 10^{-7},\,10^{-6},\text{ and }
10^{-5}\,\text{kg}\,\text{m}^{-3}$ and the velocities
$2500\text{ and }3000\,\text{km\,s}^{-1}$.
The resolution of the simulation is $2400/480$ grid cells in the radial/azimuthal direction within the quadrant.}
\label{colorfigdensenew1}
\end{center}
\end{figure}

We first study the SN-CSM interaction for the case of high density of
surrounding media. The CSM consists in the spherically symmetric component and the
aspherical circumstellar disk described in Sect.~\ref{baseq}. The parameters
of the CSM and progenitor star are given in Table~\ref{table1} and in
the first paragraph of Sect.~\ref{numero}.

Integrating the density profiles,
we estimate the total mass of the CSM components:
for spherically symmetric component of CSM we obtain the total mass 
$M_\text{sw}=4\pi R_\star^2\rho_{0,\text{sw}}\int_{R_\star}^{r_1}\d r\approx 
0.15\,M_\odot$ within the radius 
$r_1=20\,R_\star$ (within the radius of the computational domain),
while we obtain the total mass for circumstellar disk component within the same domain as
$M_\text{csd}=2\pi R_\star^2\Sigma_{0,\text{csd}}
\int_{R_\star}^{r_1}R^{-1}\dd R\approx 0.65\,M_\odot$ 
\citep[cf.][]{2011A&A...527A..84K,2014A&A...569A..23K,2018A&A...613A..75K}. 

We show in Figs.~\ref{colorfig1} and \ref{colorfig2}
the 2D snapshots of density, velocity, and temperature, in equatorial ($x$) - polar ($z$) plane, 
at selected times $t=16\,\text{h}$ (only density) and $t=66\,\text{h}$. We also add illustratively separated 
1D slices of the graphs of selected quantities in equatorial plane and in polar direction
in Fig.~\ref{fig2_all} at the time $t\approx 66\,\text{h}$, supplemented by a graph
of specific radiation entropy $s=16\sigma T^3/(3c\rho),$ 
where $\sigma$ is the Stefan-Boltzmann constant and $c$ is the speed of light 
\citep{1967pswh.book.....Z}. 
The figures show the regions of decelerated expansion velocity toward the 
equatorial disk while the expansion is unlimited (relatively, according 
to the density of the spherically symmetric CSM component) in the polar 
direction. All the characteristics show enhanced peaks in the SN-disk contact region 
and indicate shoulders and strips of overdensity and enhanced 
temperature that propagate around the equatorial disk. 

We show in the upper panel of Fig.~\ref{fig33} the density slopes $n$ and $w$ 
(see Eqs.~\eqref{homolog1} and \eqref{homolog4}) 
and in the lower panel of Fig.~\ref{fig33} the temperature slopes $p$ (cf.~Eq.~\eqref{temperature}),
calculated in the numerical model at the same time $t\approx 66\,\text{h}$ 
since shock emergence. However, the outer shock, which roughly separates the outer region 
(denoted in Sect.~\ref{sedov} by subscript ``out'') from all the other regions between 
the two shock waves, moves to the radius approximately 
$11\,R_\star$ in the equatorial plane and $16.5\,R_\star$ in the polar direction.
The inner envelope density slope ranges in the polar direction between $n=0$ 
and $n\approx 9,$ while in the equatorial plane the density slope is highly discontinuous 
and becomes much steeper. However, we cannot simply apply the analytical 
similarity relations in the equatorial plane with respect to the constraint $w <3$ given by Eq.~\eqref{aaas7a}.
The temperature slope in the smooth region 
of the inner envelope (excluding the areas of sharp discontinuities) 
ranges between $p=0$ and $p\approx 2.7$ in the equatorial and in the polar direction.
In order to avoid inadequate computational difficulties, 
we calculate the temperature in all the domain as radiative 
 ($T\sim P^{\,0.25}$; the same applies for the other models). 
 We may regard the displayed distances reached by the forward shock wave 
 in the equatorial and in the polar direction as
 a certain scale of the level of asphericity (except the slopes of the selected quantities) 
 of the SN explosion; the same also applies for another models (see Figs.~\ref{fig33_mid} and ~\ref{fig33_sparse}).

In Figs.~\ref{fig44} and \ref{fig55} we demonstrate the semi-analytically found (see Sect.~\ref{semianal}) 
similarity solution of density, velocity, pressure, and temperature for two inner density slopes 
$n=7$ and $n=9$ that correspond to the average and maximum numerically 
calculated inner density slopes in the polar direction where
the similarity solution holds and where the parameter $w<3$ 
(see Eq.~\eqref{homolog4} and its consequences). In the disk-residing 
equatorial plane the analytical similarity solution is 
however violated because of the disk equatorial plane density profile $\rho_\text{eq}(R)\sim R^{-3.5}$ ($w>3$).

Comparing slices of 2D numerical calculations in Fig.~\ref{fig2_all} 
with the 1D semi-analytical solution (cf.~Sect.~\ref{semianal}) in polar direction in
Figs.~\ref{fig44} and \ref{fig55},
we find that the solutions well agree in global parameters. That is,
density $\rho$ decreases from the value $\rho\approx 5\times 10^{-6}\,\text{kg}\,\text{m}^{-3}$ 
at the radius $r=11.5\,R_\star$ to 
$\rho\approx 3\times 10^{-7}\,\text{kg}\,\text{m}^{-3}$
behind the outer shock, while it drops to almost 
$10^{-8}\,\text{kg}\,\text{m}^{-3}$ outside the outer shock; the radial (expansion)
velocity $u$ grows in the same domain from $u\approx 3\,000\,\text{km}\,\text{s}^{-1}$ to 
approximately $3\,500\,\text{km}\,\text{s}^{-1}$ behind the shock 
wave zone. We hereafter denote the radial velocity of expansion $u$ for
simplicity; its value in fact does not much differ 
from the total magnitude of velocity since the nonradial velocity 
components are more than an order of magnitude smaller.  
The pressure approaches $P\approx 10^6\,\text{Pa,}$ while 
the corresponding temperature $T$ is slightly higher than $10^5\,\text{K}$ 
(compare Figs.~\ref{fig44} and \ref{fig55} with Fig.~\ref{fig2_all}). 
The semi-analytical solution is directly compared
with 1D slices of our 2D numerical solution for the same time of calculation. We also add the
1D synchronous profile
calculated using the SNEC-1.01 code \citep{2015ApJ...814...63M}, 
which takes into account 
more realistic shape of the density enhancement zone between reverse shock and contact discontinuity 
(between $13.5$ and $13.8\, r/R_\star$) that is
formed during propagation of the shock wave through the stellar body \citep{1959sdmm.book.....S,Sakurai}.
Since our current models use the whole stellar body as an initial thermal bomb, the shock wave directly propagates 
 (like a Riemann-Sod shock wave) to the region of dropped density. 
 However, this detail in the shock wave structure 
 has only a minor impact on the global solution within the study.
For the SNEC calculation we input the same initial profile of the progenitor star and 
the CSM as for all the other calculations with radiation included,
the initial thermal bomb zone \citep{2015ApJ...814...63M} 
is however limited only to a relatively small region near the stellar center.

We also constructed similar comparative
diagrams for other values of parameter $n$ lower than $n=7$. 
However, in this case the radial distance between the inner and outer shock waves becomes
too large and the values of analytically calculated quantities differ significantly from SNEC and from our 2D numerical result. In any case, we found that the analytical versus numerical solutions 
are closest (relatively) at this stage
for the internal density slope parameter $n\approx 7$ - $n\approx 9$. 

We also checked the constraints on the applicability 
of the adiabatic self-similar solution within the semi-analytical calculation (see Appendix \ref{limanal}). 
For the model with high density (Sect.~\ref{numerodense})
and the density slope $n=7$ we found the upper limit $t_\text{max}\approx 20\,\text{yrs,}$ 
while the lower limit $t_\text{min}$ is on the order 
of several minutes at the start of the calculation. The lower limit continuously grows during the time evolution,
however, the values of $t_\text{min}$ remain several orders 
of magnitude lower than the actual time of simulation.

Fig.~\ref{colorfig2} indicates that the zone of shear between the slower expansion into the 
region of circumstellar equatorial disk, where the vertical motion occurs
owing to the violation of disk hydrostatic equilibrium caused by the interaction of the two masses. Figure 3 also shows that the spherically symmetric fast SN expansion may provoke the development
of Kelvin-Helmholtz instabilities.
This is further demonstrated in Fig.~\ref{colorfigdensenew1}, where we show
the snapshot of 2D density and expansion velocity field in later time $t\approx 165\,\text{h}$ (up to the radius
$50\,R_\star$) and in higher numerical resolution. We thus get 
a comparison with the model in Fig.~\ref{colorfig2}, where we can distinguish
the details of evolution of the selected quantities in yet more evolved and more structured stage. 
The shoulders of overdensity and significantly decelerated expansion velocity at the zone of shear between the disk and spherically symmetric CSM are illustrated in a more detailed
picture. There is also a clear indication of developing Kelvin-Helmholtz and Rayleigh-Taylor instabilities along the zone of shear 
interaction and SN-disk contact region.
\begin{figure} [t!]
\begin{center}
\centering\resizebox{0.725\hsize}{!}{\includegraphics{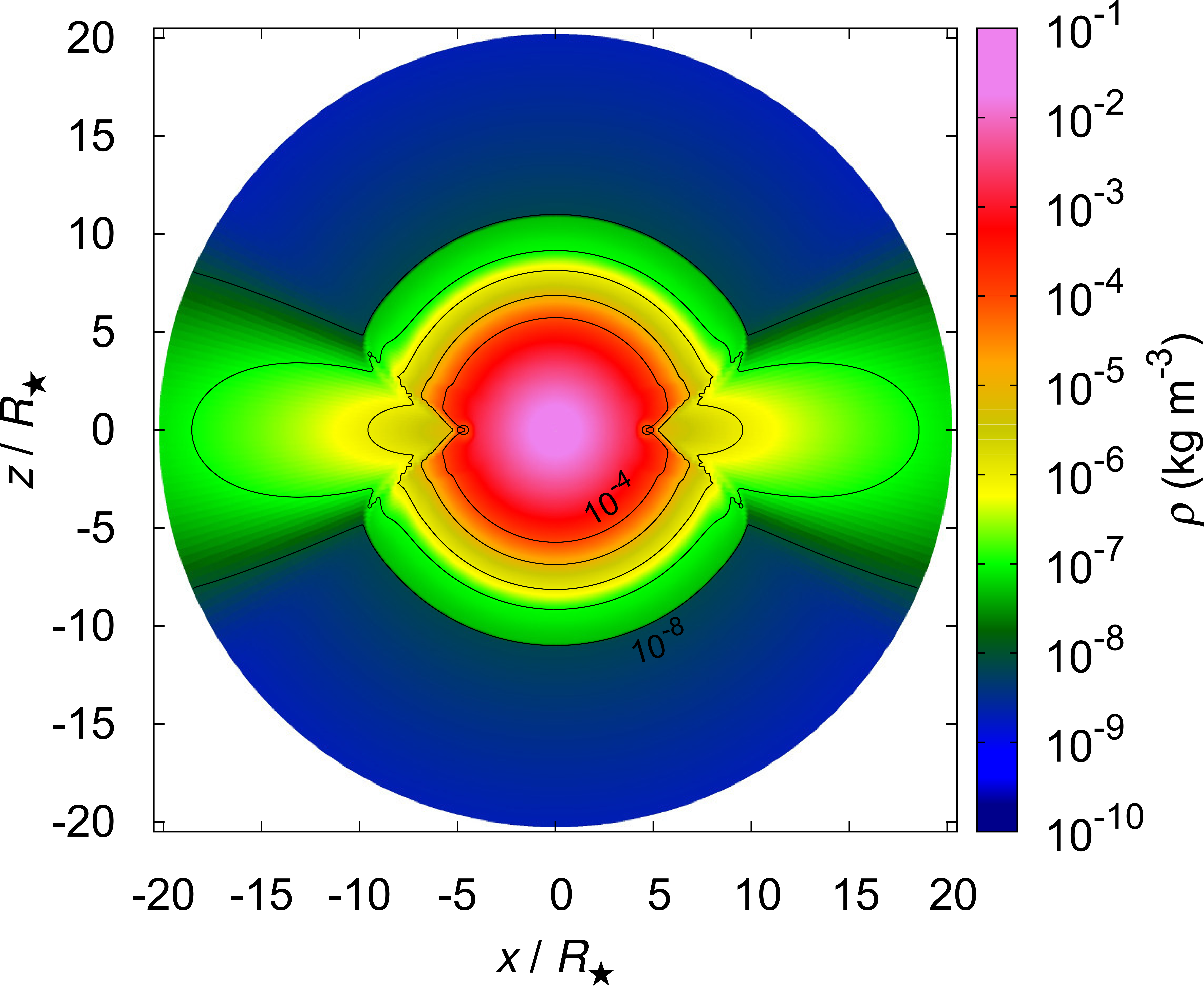}} 
\centering\resizebox{0.725\hsize}{!}{\includegraphics{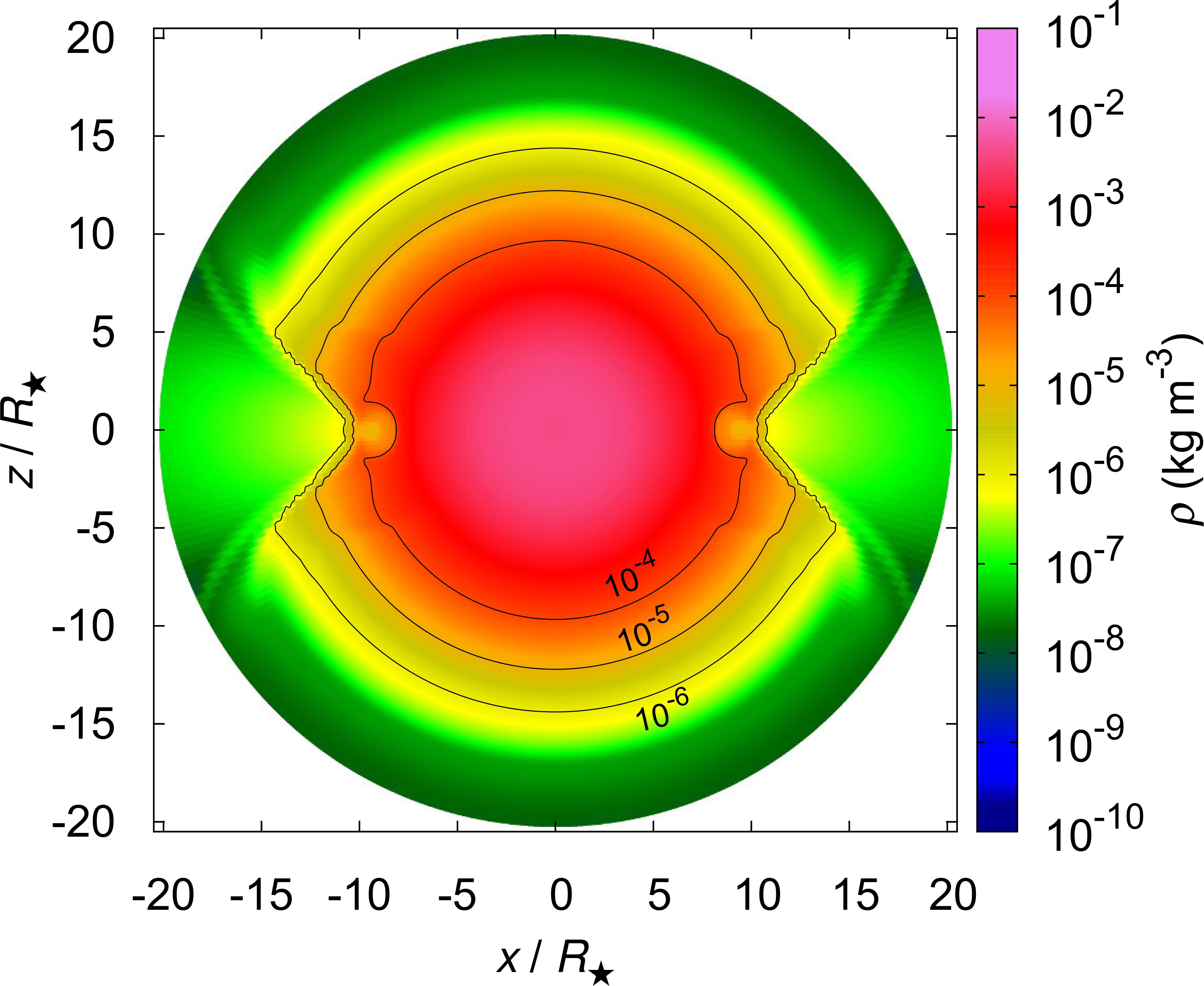}} 
\caption{Model B: \textit{Upper panel}: Color map of the density structure 
at time $t \approx 33\,\text{h}$ since shock emergence.
Contours denote the densities $\rho = 10^{-8},\,10^{-7},\,10^{-6},\,10^{-5},\text{ and }
10^{-4}\,\text{kg}\,\text{m}^{-3}$.
Characteristic 1D sections of the model in the equatorial and polar plane are shown in Fig.~\ref{fig4_all}.
\textit{Lower panel}: Color map of the density structure 
at time $t \approx 66\,\text{h}$ since shock emergence.
Contours denote the densities $\rho = 10^{-6},\,10^{-5},\text{ and }10^{-4}\,\text{kg}\,\text{m}^{-3}$.}
\label{colorfig11}
\end{center}
\end{figure}
\subsection{Model B: Intermediate density}\label{numeromid}
\begin{figure} [t!]
\begin{center}
\centering\resizebox{0.725\hsize}{!}{\includegraphics{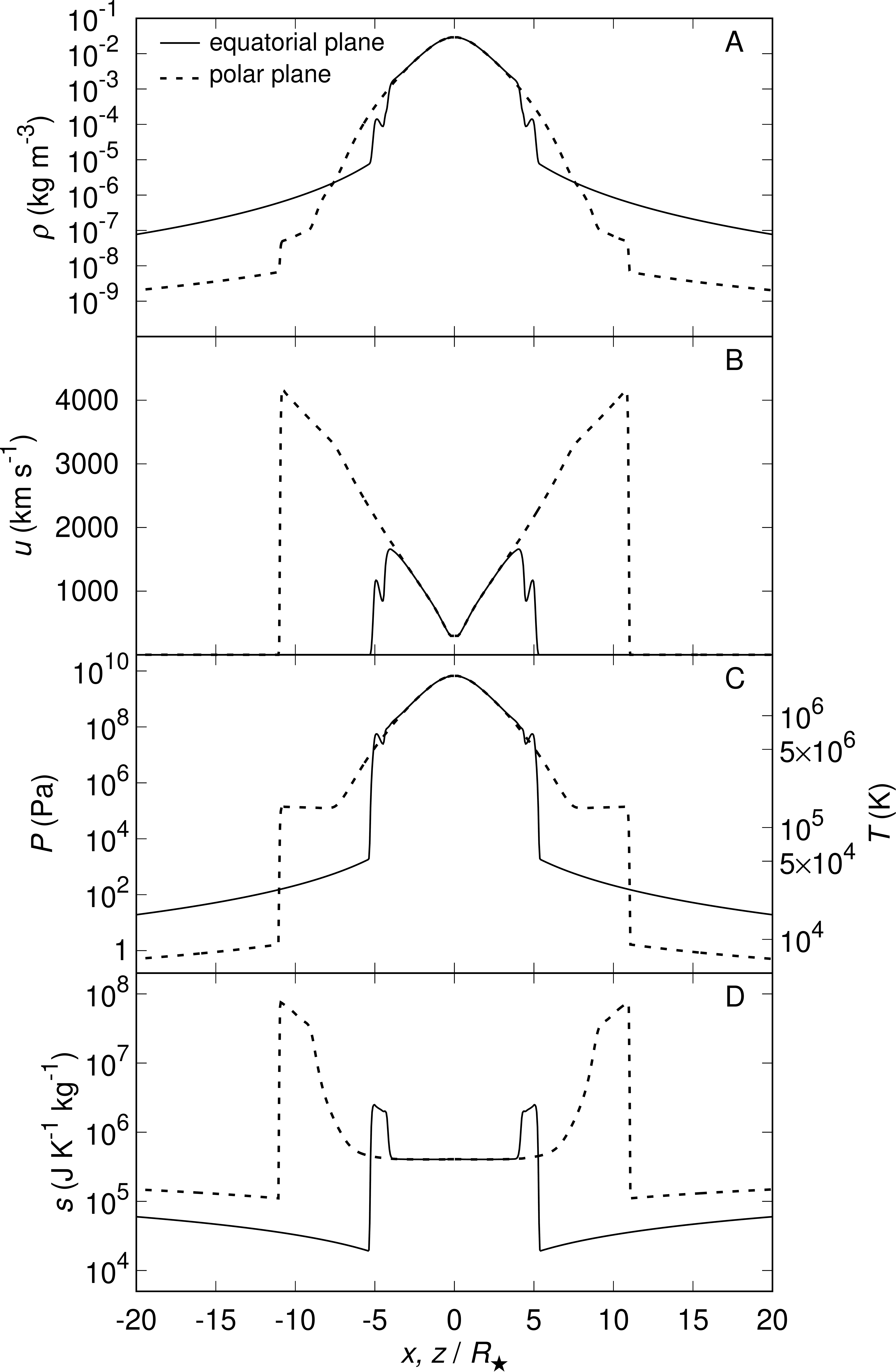}} 
\caption{Model B: 1D slices of hydrodynamical variables
in the equatorial plane ($x$-coordinate, solid line) and polar plane ($z$-coordinate, dashed line)
at time $t \approx 33\,\text{h}$ since shock emergence (as in Fig.~\ref{colorfig11}).
\textit{Panel A}: The density $\rho$.
\textit{Panel B}: The expansion velocity $u$.
\textit{Panel C}: Pressure $P$ and temperature $T$.
\textit{Panel D}: Specific entropy $s$.}
\label{fig4_all}
\end{center}
\end{figure}
\begin{figure} [t!]
\begin{center}
 \centering\resizebox{0.65\hsize}{!}{\includegraphics{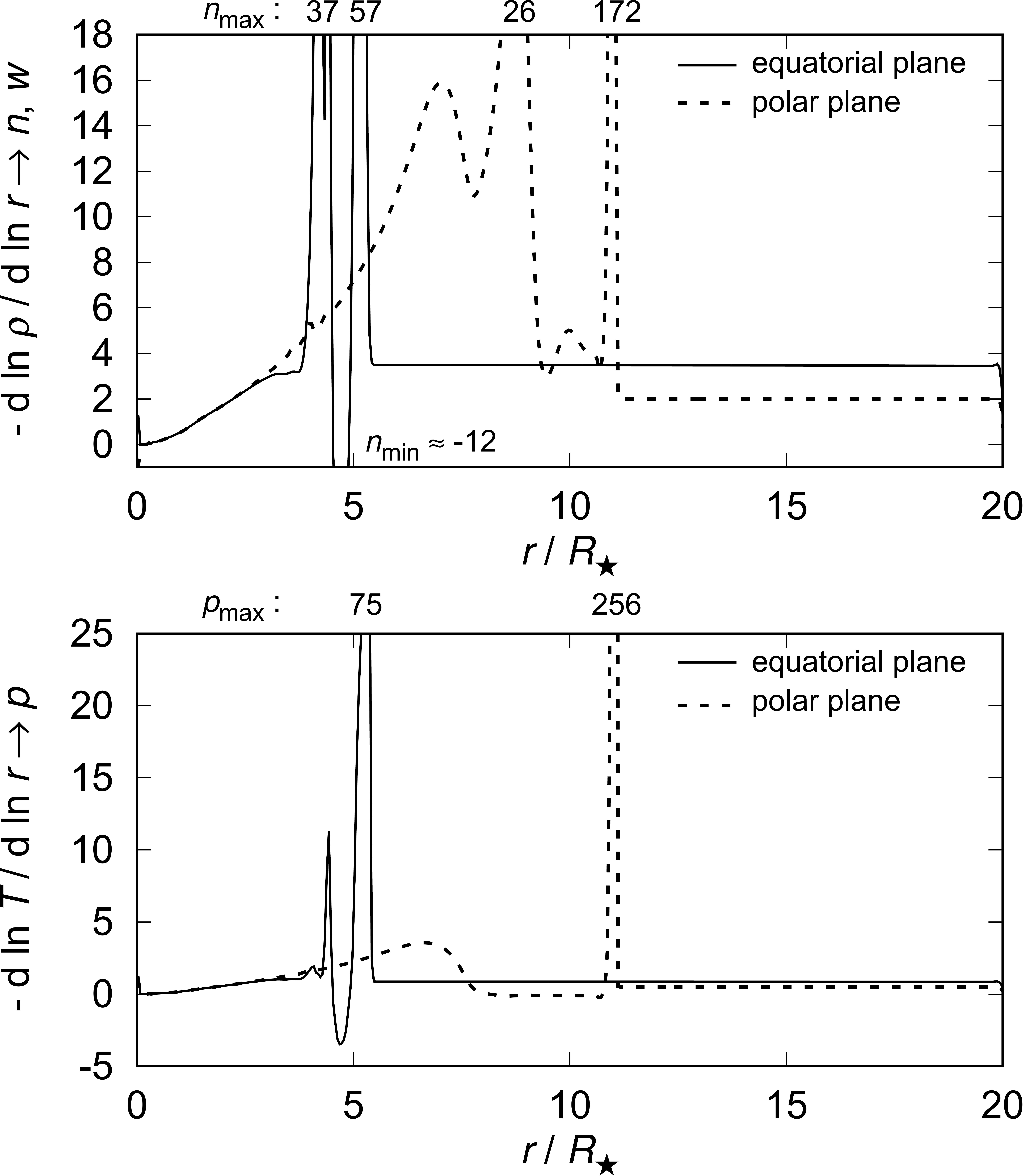}} 
 \caption{Model B: \textit{Upper panel}: 
 Slope parameters $n$ and $w$ of the density (analogous to Fig.~\ref{fig33}) 
 in the equatorial (solid line) and polar (dashed line) direction, corresponding to
 panel A in Fig.~\ref{fig4_all} at the same time $t\approx 33\,\text{h}$. 
 The position of the outer shock wave is at 
 $r\approx 5.5\,R_\star$ in the equatorial direction, while it is at $r\approx 11\,R_\star$ in the polar direction.
 The inner envelope density slope parameter increases from $n=0$ to $n\approx 16$ in the polar direction. 
 {\textit{Lower panel}: Slope parameters $p$ of the temperature in the equatorial (solid line) 
 and polar (dashed line) direction
 corresponding to the temperature graph (panel C) in Fig.~\ref{fig4_all} at the same time. 
 The inner envelope (smooth) temperature slope parameter increases to $p\approx 1.9$ at $r\approx 4.1,R_\star$ in
 the equatorial direction while 
 it increases to $p\approx 3.6$ at $r\approx 6.6\,R_\star$ in the polar direction.}}
 \label{fig33_mid}
 \end{center}
\end{figure}
\begin{figure} [t!]
\begin{center}
\centering\resizebox{0.85\hsize}{!}{\includegraphics{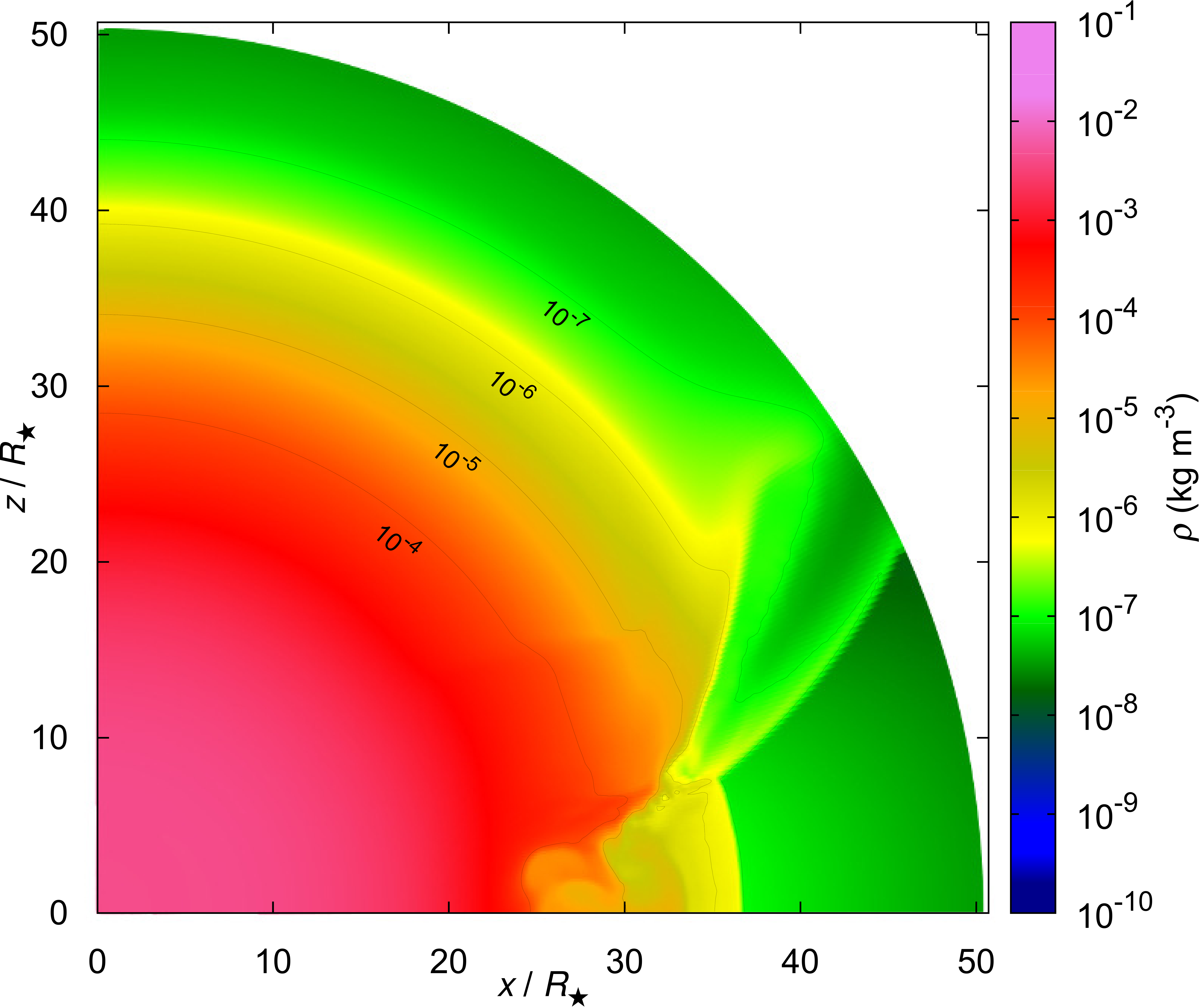}} 
\centering\resizebox{0.85\hsize}{!}{\includegraphics{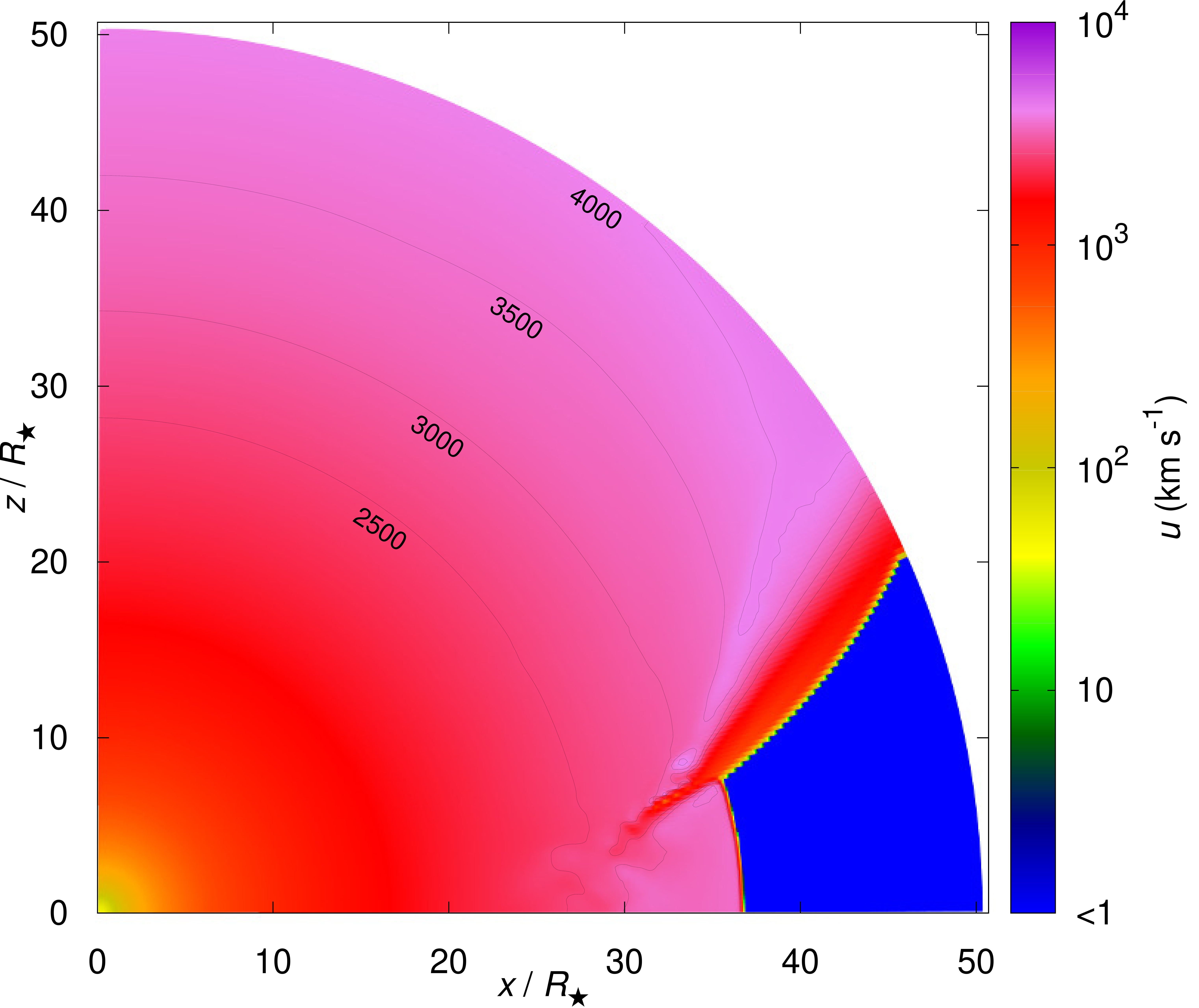}} 
\centering\resizebox{0.85\hsize}{!}{\includegraphics{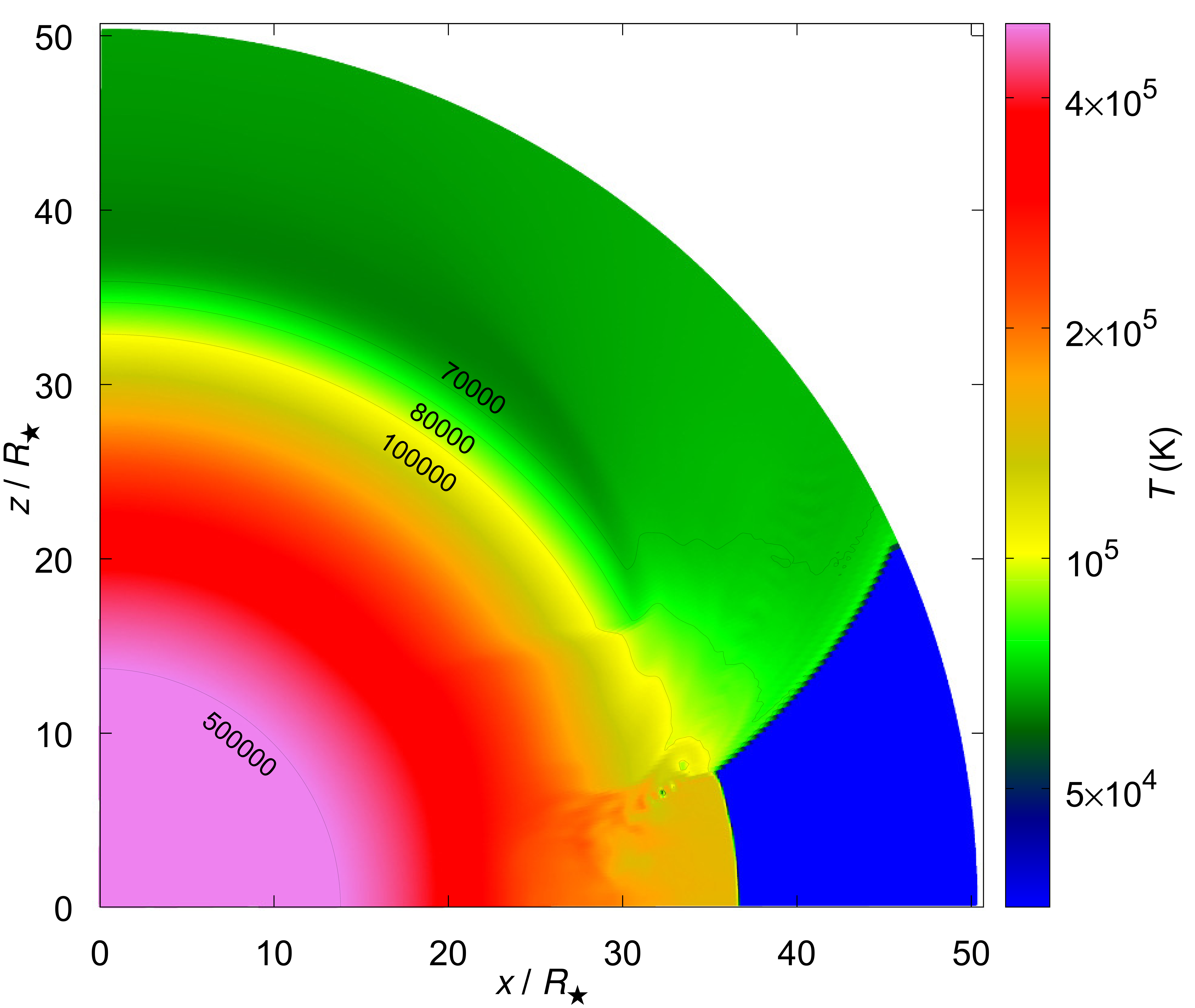}} 
\caption{Model B:
Detailed color maps of the density (\textit{upper panel}),
radial velocity (\textit{middle panel}), and temperature (\textit{lower panel}) 
structure up to the distance $50\,R_\star$ at time $t \approx 180\,\text{h}$ 
since shock emergence, with the same numerical resolution as in Fig.~\ref{colorfigdensenew1}.
Contours denote the densities $\rho = 10^{-7},\,10^{-6},\,10^{-5},\text{ and }
10^{-4}\,\text{kg}\,\text{m}^{-3}$,
the velocities $2500,\,3000,\,3500,\text{ and }4000\,\text{km\,s}^{-1}$,
and the temperatures $7\times 10^{4},\,8\times 10^{4},\,10^{5},\text{ and }5\times 10^{5}\,\text{K}$.}
\label{colorfigmidnew1}
\end{center}
\end{figure}
\begin{figure} [t!]
\begin{center}
\centering\resizebox{0.75\hsize}{!}{\includegraphics{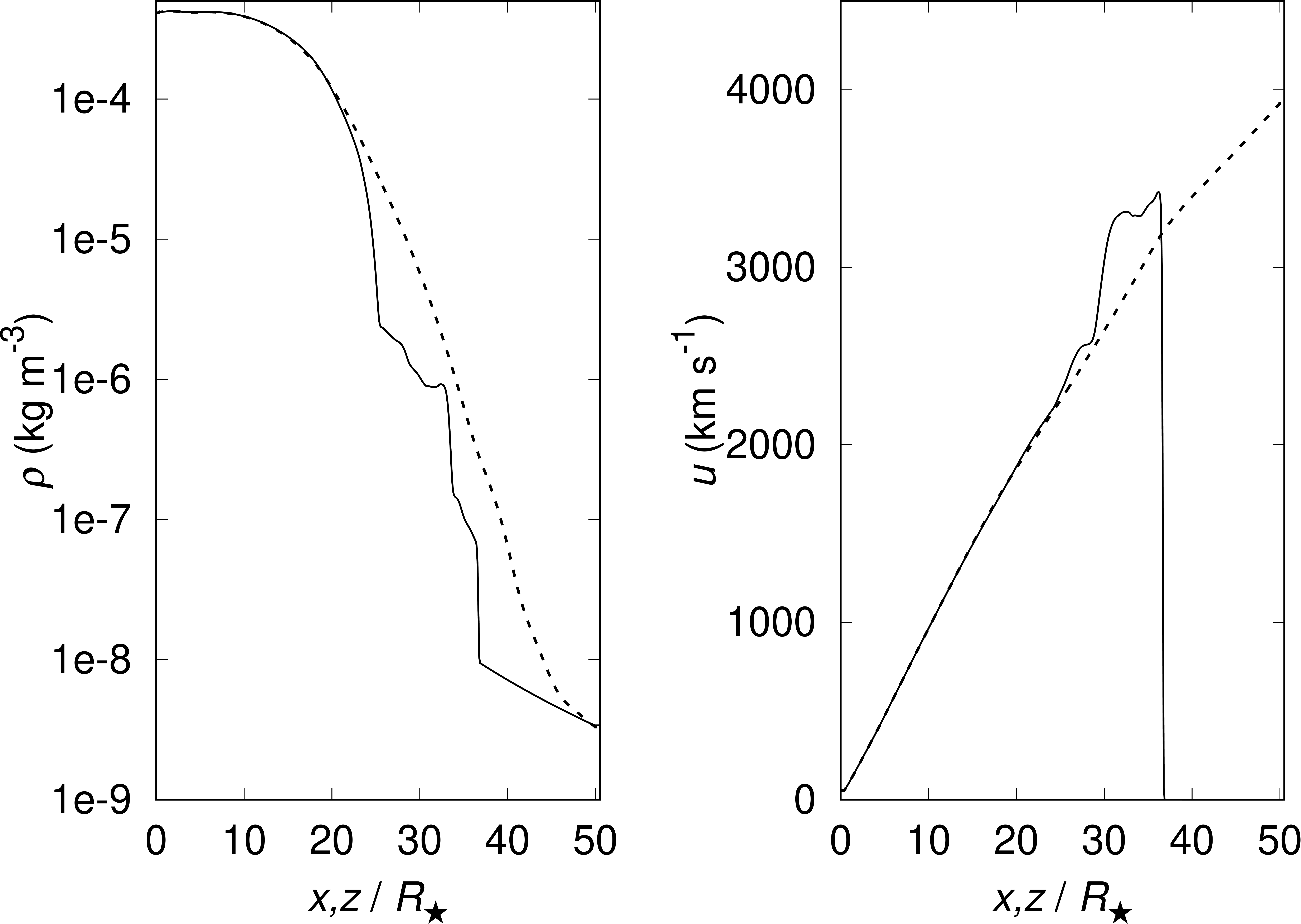}} 
\caption{Model B:
1D slices of density (\textit{left panel}) and velocity (\textit{right panel}) in the equatorial plane
($x$-coordinate, solid line) and in polar direction ($z$-coordinate, dashed line)
at time $t\approx 180\,\text{h}$ since shock emergence 
(corresponding to Fig.~\ref{colorfigmidnew1}). 
The maximum of the polar velocity, which is already outside the computational domain, is
approximately $4400\,\text{km}\,\text{s}^{-1}.$}
\label{colorfig1Dmidnew1}
\end{center}
\end{figure}
\begin{figure} [t!]
\begin{center}
\centering\resizebox{0.75\hsize}{!}{\includegraphics{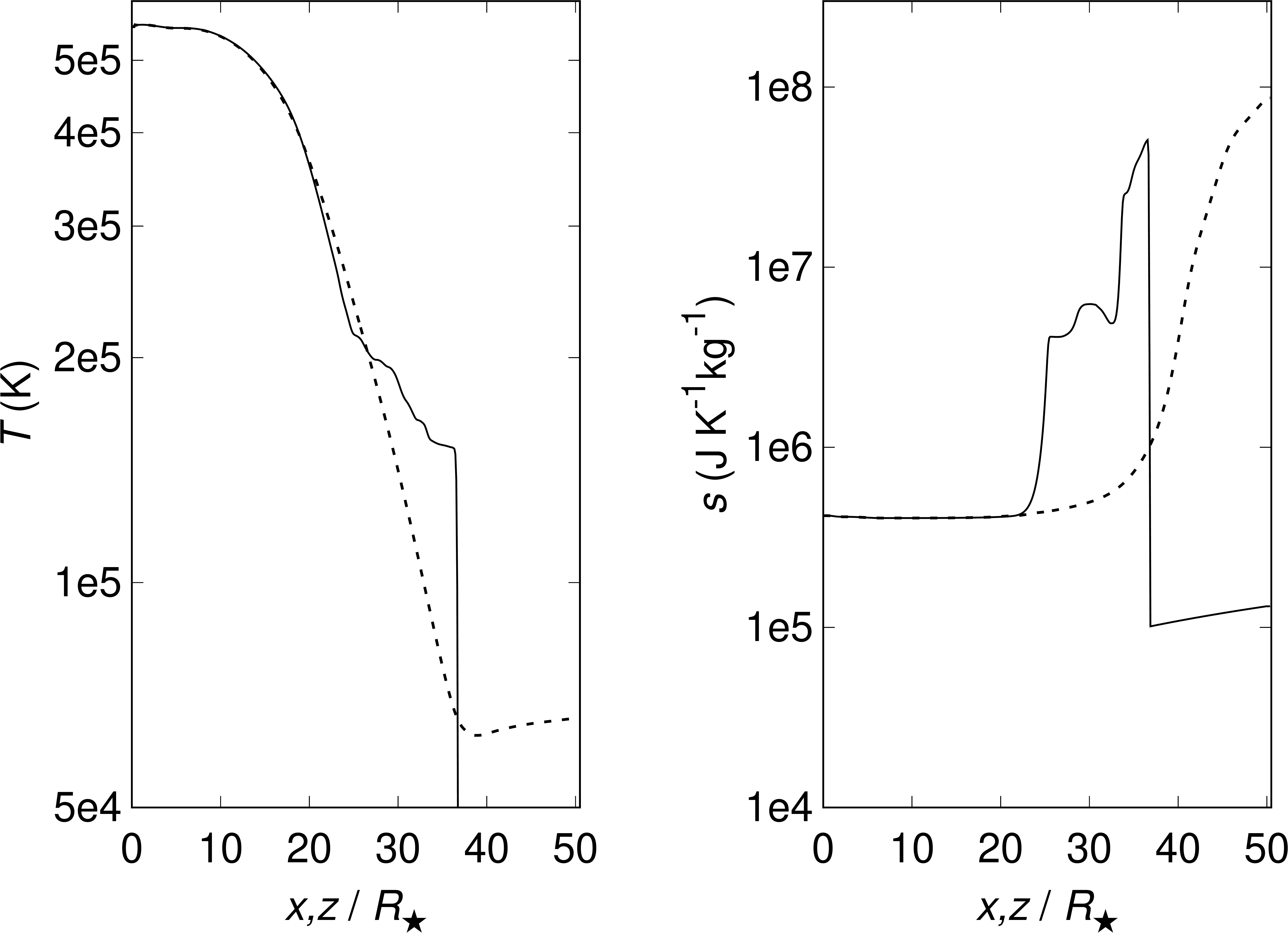}} 
\caption{Model B:
1D slices of temperature (\textit{left panel}) and specific entropy (\textit{right panel}) in the equatorial
($x$-coordinate, solid line) and in polar direction ($z$-coordinate, dashed line)
at the same time as in Fig.~\ref{colorfig1Dmidnew1}.}
\label{colorfig1Dmidnew2}
\end{center}
\end{figure}
We further studied interaction of an SN blast wave with the CSM for the case of
intermediate density (Model B).
The progenitor and CSM parameters are introduced in the first paragraph 
of Sect.~\ref{numero} and in Tab.~\ref{table1}.
Integrating the CSM density profiles,
we obtain the following total mass of the CSM components:
the total mass $M_\text{sw}\approx 0.015\,M_\odot$ 
within the radius of the computational domain $r_1=20\,R_\star$ and
$M_\text{csd}\approx 0.65\,M_\odot$ 
\citep[cf.][]{2011A&A...527A..84K,2014A&A...569A..23K,2018A&A...613A..75K} 
within the same domain. 

We show (analogous to Sect.~\ref{numerodense}) in Fig.~\ref{colorfig11}
the 2D snapshots of density profiles in equatorial ($x$) - polar ($z$) plane 
at selected times $t=33\,\text{h}$ and $t=66\,\text{h}$, respectively. 
The lower panel in Fig.~\ref{colorfig11} shows significant layers of overdensity that are formed
near the zone where the expanding matter is forced to move along 
the interface between the spherically symmetric component of CSM 
and the denser circumstellar equatorial disk region of CSM.
The development of Kelvin-Helmholtz instabilities is also remarkably indicated near the shear zones.
We also add in Fig.~\ref{fig4_all} the separated 
1D slices of graphs of density, velocity, pressure, temperature, 
and specific radiation entropy in equatorial and polar plane at the time $t=33\,\text{h}$.
The model shows higher difference in expansion velocity between the equatorial and 
polar direction than in the previous model because of the lower density (lower mass injection) of 
spherically symmetric CSM. The front of the polar SN expansion reaches the
velocity approximately $4000\,\text{km}\,\text{s}^{-1}$ while it approaches 
$2000\,\text{km}\,\text{s}^{-1}$ in the equatorial direction.
Because the expansion shock front in the polar direction is
too close to the limit of computational area in the time for which we
plotted the previous models ($t\approx 66\,\text{h}$), we provide for
this model the profiles of most of the characteristics in earlier time ($t\approx 33\,\text{h}$) 
while, on the other hand, 
the graphs of the previous model with high density would be in this time poorly illustrative. 

We also show (analogously to Sect.~\ref{numerodense}) in the upper
panel of Fig.~\ref{fig33_mid} the density slopes $n$ and $w$,
while in the lower panel we show the temperature slopes $p$,
calculated at the same time $t\approx 33\,\text{h}$ since shock emergence. 
The outer shock moves at the radius of approximately 
$5.5\,R_\star$ in the equatorial and 
$11\,R_\star$ in the polar direction. 
The inner envelope density slope ranges in the polar direction between 
$n=0$ and $n\approx 16$, the temperature slope 
in the smooth region of the inner envelope 
ranges between $p=0$ and $p\approx 1.9$ in the equatorial direction and between 
$p=0$ and $p\approx 3.6$ in the polar direction.

The calculation of the constraints of applicability of the adiabatic self-similar solution 
(see Appendix \ref{limanal}) gives
for the model with intermediate density the upper limit 
$t_\text{max}\approx 21\,\text{yrs}$ for the density slope $n=7,$ while 
for $n=12$ it increases to  $t_\text{max}\approx 35\,\text{yrs}$. 
For the lower limit $t_\text{min}$ applies the same as in previous model 
with high density in Sect.~\ref{numerodense}.

Analogous to the previous model, we show in Fig.~\ref{colorfigmidnew1} the snapshots of 2D density, 
radial expansion velocity, and temperature
in time $t\approx 180\,\text{h}$ since shock emergence, 
up to the radius $50\,R_\star$, in high numerical resolution, 
performed on 2D grid with 2400 grid cells in radial and 480 grid cells in azimuthal direction. 
Comparing this with Fig.~\ref{colorfig11}, we distinguish the details of evolution of the selected 
quantities, including the even more significant shoulders of overdensity and over-heated gas 
near disk - spherically symmetric CSM shear zone. 
Although the forward shock front is in the snapshot already out of the computational domain, it particularly well 
illustrates the development of Kelvin-Helmholtz and Rayleigh-Taylor instabilities which is clearly indicated.

We also add in this intermediate density model the 1D equatorial, polar slices
of density and expansion velocity (Fig.~\ref{colorfig1Dmidnew1}), and the temperature and specific entropy 
(Fig.~\ref{colorfig1Dmidnew2}), which correspond to the instant time shown in Fig.~\ref{colorfigmidnew1}. 
A qualitative comparison with the profiles in Fig.~\ref{fig4_all} (in less advanced time) shows the decrease of density, 
however in a (roughly) self-similar mode.
This is accompanied by an increase of the equatorial velocity peak.
The peak of the polar velocity is currently already outside the computational
domain, nevertheless the results of the calculation before it reaches the outer
boundary show that the polar velocity increases very little (cf.~the
increase of polar velocity in the model with high density (Sect.~\ref{numerodense}) 
by comparing  the values of the polar velocity maximum in 
Figs.~\ref{colorfig2} and \ref{fig2_all} with Fig.~\ref{colorfigdensenew1}).

\subsection{Model C: Low density}\label{numeroweak}
\begin{figure} [t!]
\begin{center}
\centering\resizebox{0.725\hsize}{!}{\includegraphics{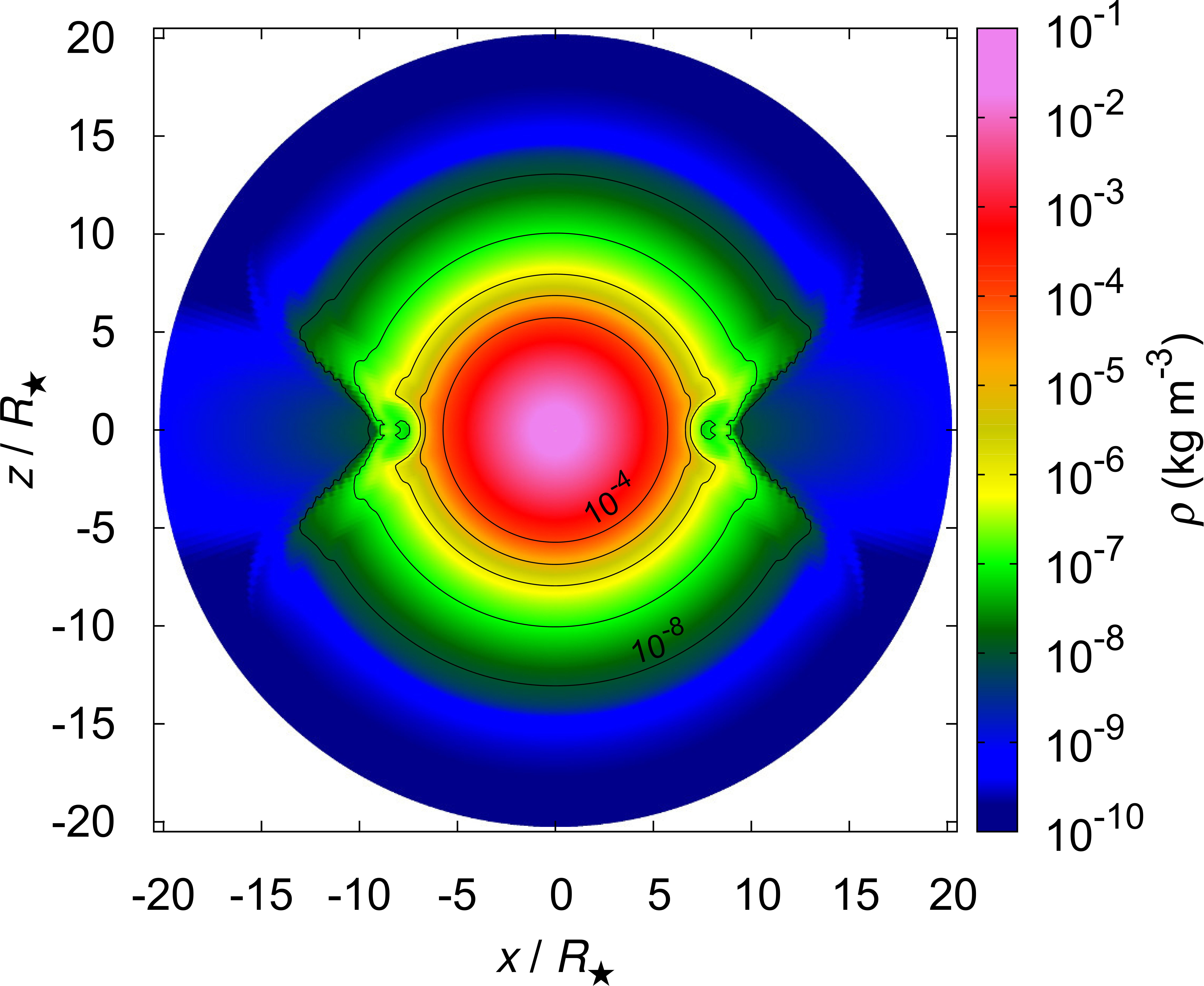}} 
\centering\resizebox{0.725\hsize}{!}{\includegraphics{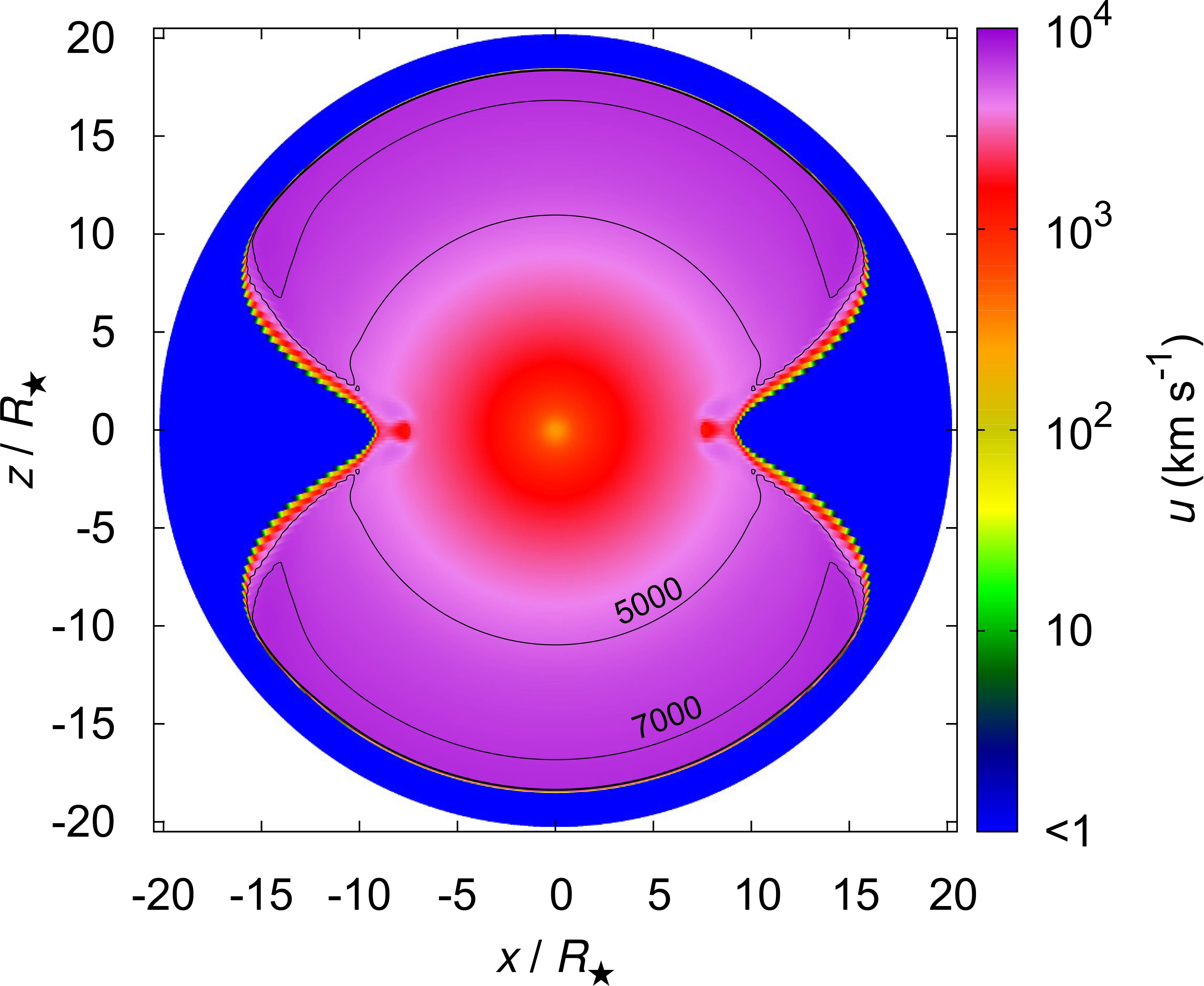}} 
\centering\resizebox{0.725\hsize}{!}{\includegraphics{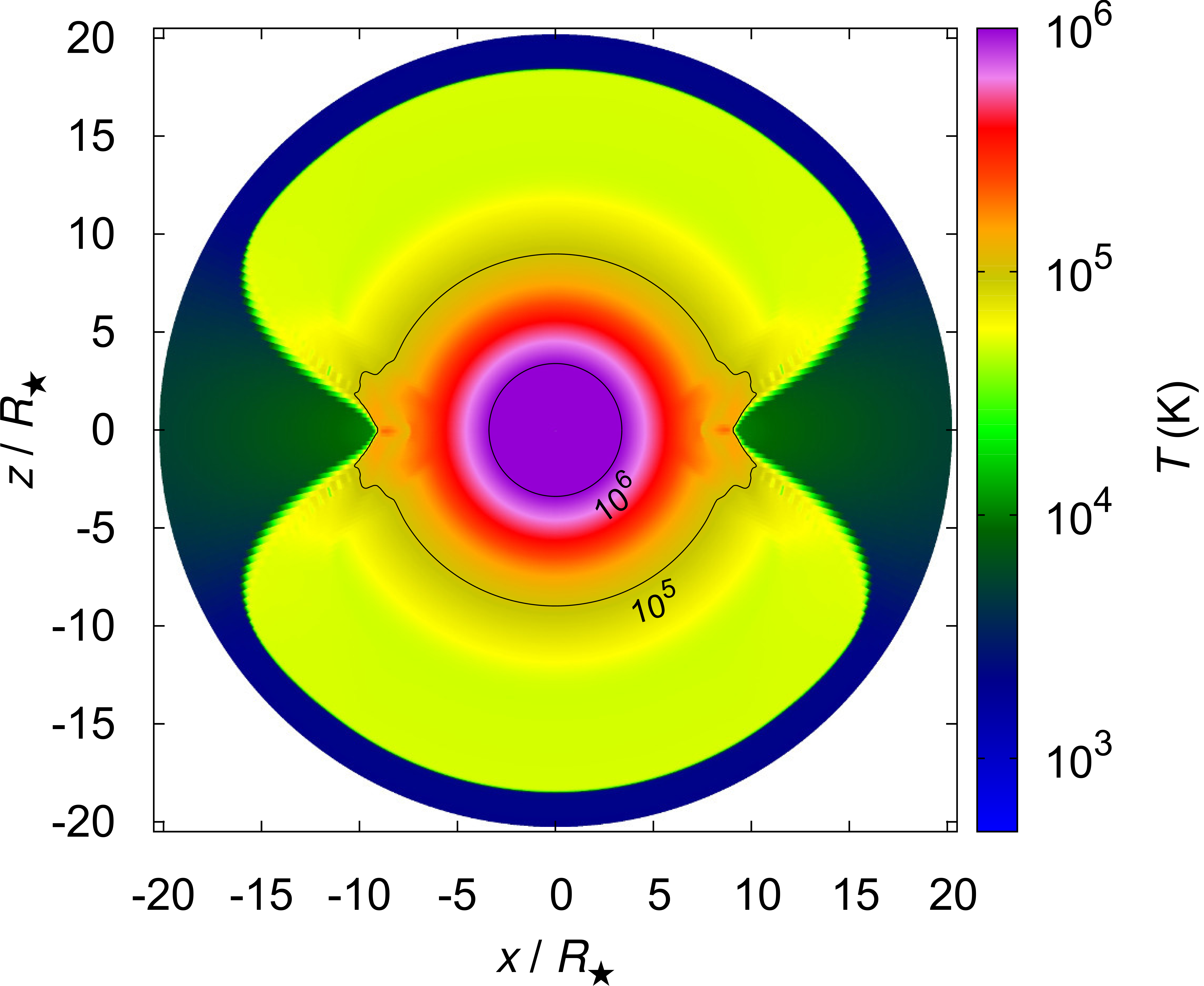}} 
\caption{Model C:
Color maps of the hydrodynamical variables
at time $t \approx 33\,\text{h}$ since shock emergence.
\textit{Upper panel}: Color map of the density structure.
Contours denote the densities $\rho = 10^{-8},\,10^{-7},\,10^{-6},\,10^{-5},\text{ and }
10^{-4}\,\text{kg}\,\text{m}^{-3}$.
\textit{Middle panel}: Color map of the velocity structure.
Contours denote the velocities 
$u = 5000\text{ and }7000\,\text{km}\,\text{s}^{-1}$.
\textit{Lower panel}: Color map of the temperature structure.
Contours denote the temperatures $T = 10^{5}\text{ and }10^{6}\,\text{K}$.
Characteristic 1D sections of the model in the equatorial
plane and in the polar direction
are shown in Fig.~\ref{fig6_all}.}
\label{colorfig21}
\end{center}
\end{figure}
\begin{figure} [t!]
\begin{center}
\centering\resizebox{0.725\hsize}{!}{\includegraphics{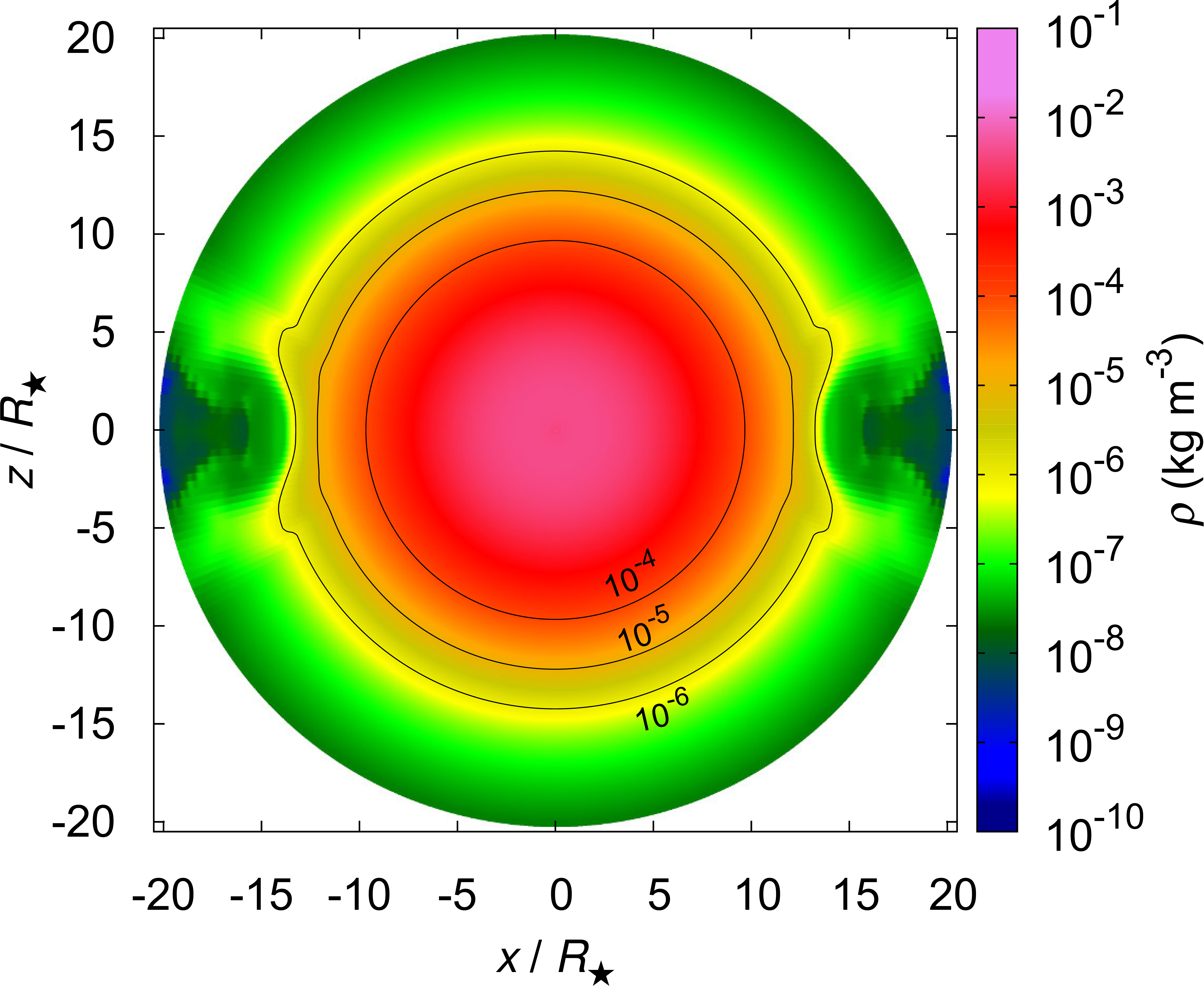}} 
\caption{Model C:
Color map of the density structure of the interaction 
at time $t \approx 66\,\text{h}$ since shock emergence.
Contours denote the densities 
$\rho = 10^{-6},\,10^{-5},\text{ and }10^{-4}\,\text{kg}\,\text{m}^{-3}$.}
\label{colorfig22}
\end{center}
\end{figure}
\begin{figure} [t!]
\begin{center}
\centering\resizebox{0.75\hsize}{!}{\includegraphics{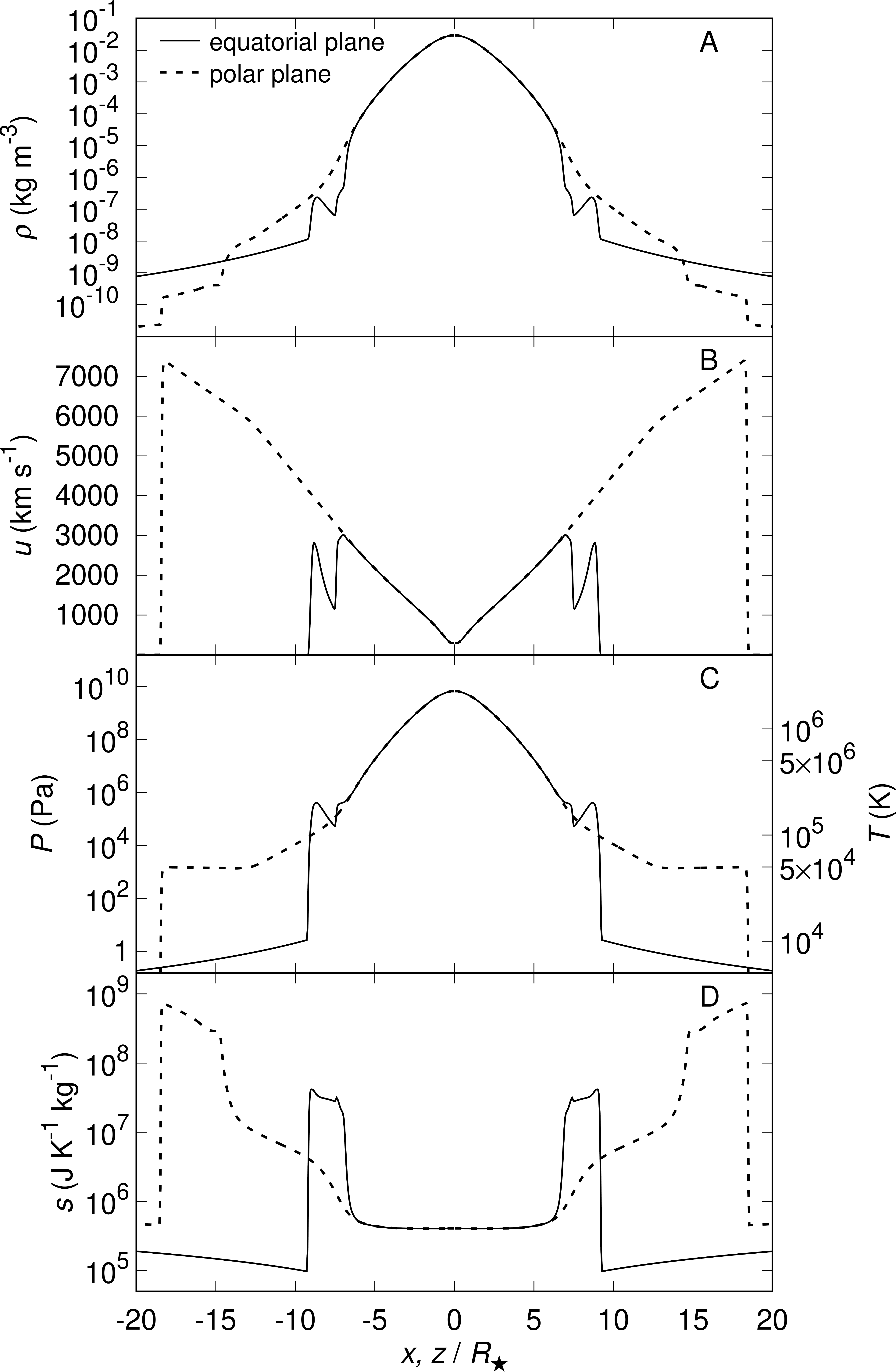}} 
\caption{Model C:
1D slices of hydrodynamical variables in the equatorial ($x$-coordinate, solid line) and 
in polar direction ($z$-coordinate, dashed line)
at time $t \approx 33\,\text{h}$ since shock emergence (as in Fig.~\ref{colorfig21}).
\textit{Panel A}: The density $\rho$. \textit{Panel B}: The expansion velocity
$u$. \textit{Panel C}: Pressure $P$ and temperature $T$. \textit{Panel D}: The
specific entropy $s$.}
\label{fig6_all}
\end{center}
\end{figure}
\begin{figure} [t!]
\begin{center}
 \centering\resizebox{0.725\hsize}{!}{\includegraphics{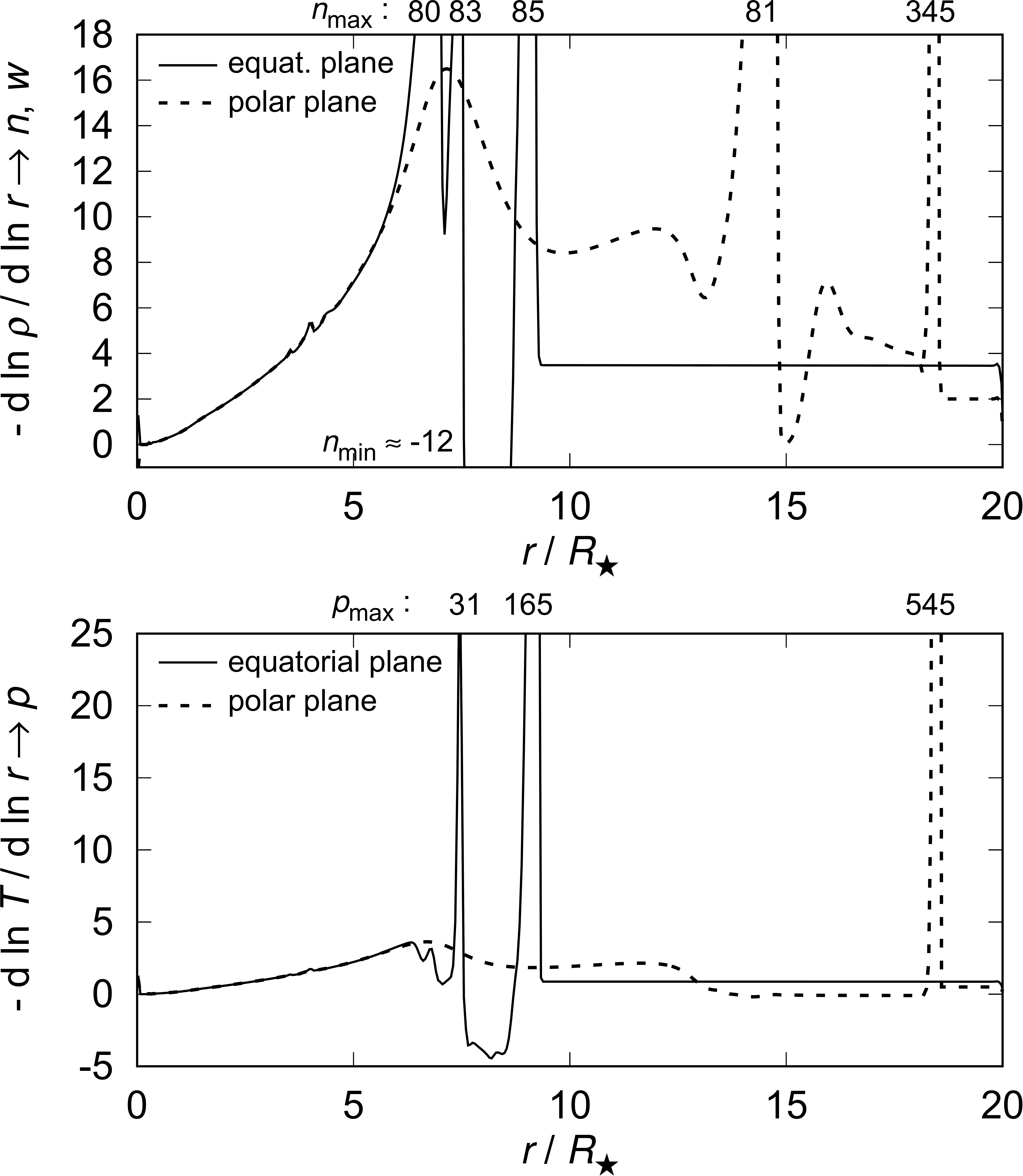}} 
  \caption{Model C:
  \textit{Upper panel}: 
 Slope parameters $n$ and $w$ of the density (analogous to Figs.~\ref{fig33} and \ref{fig33_mid}) 
 in the equatorial (solid line) and polar (dashed line) direction, 
 corresponding to panel A in Fig.~\ref{fig6_all} at the same time $t\approx 33\,\text{h}$. 
 The position of the outer shock wave is at 
 $r\approx 9\,R_\star$ in the equatorial direction while it is at $r\approx 18\,R_\star$ in the polar direction.
 The inner envelope density slope parameter increases from $n=0$ to $n\approx 17$ in the polar direction. 
 {\textit{Lower panel}: Slope parameters $p$ of the temperature in the equatorial 
 (solid line) and polar (dashed line) direction
 corresponding to the temperature graph (panel C) in Fig.~\ref{fig6_all} at the same time. 
 The inner envelope (smooth) temperature slope parameter increases to 
 $p\approx 3.6$ at $r\approx 6.3\text{~-~}6.5\,R_\star$ in
 the equatorial and also in the polar direction.}}
 \label{fig33_sparse}
 \end{center}
\end{figure}
\begin{figure} [t!]
\begin{center}
\centering\resizebox{0.85\hsize}{!}{\includegraphics{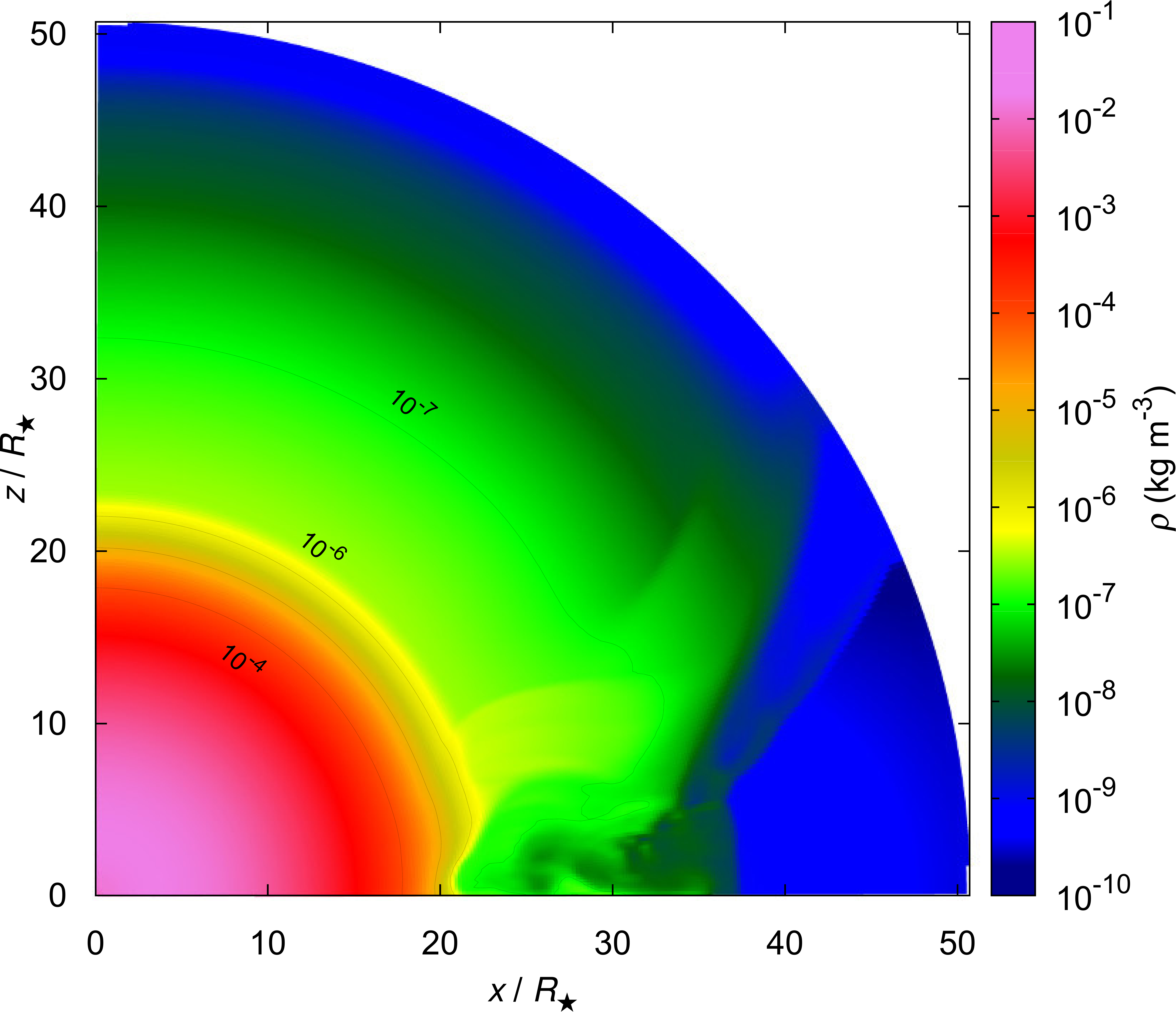}} 
\centering\resizebox{0.85\hsize}{!}{\includegraphics{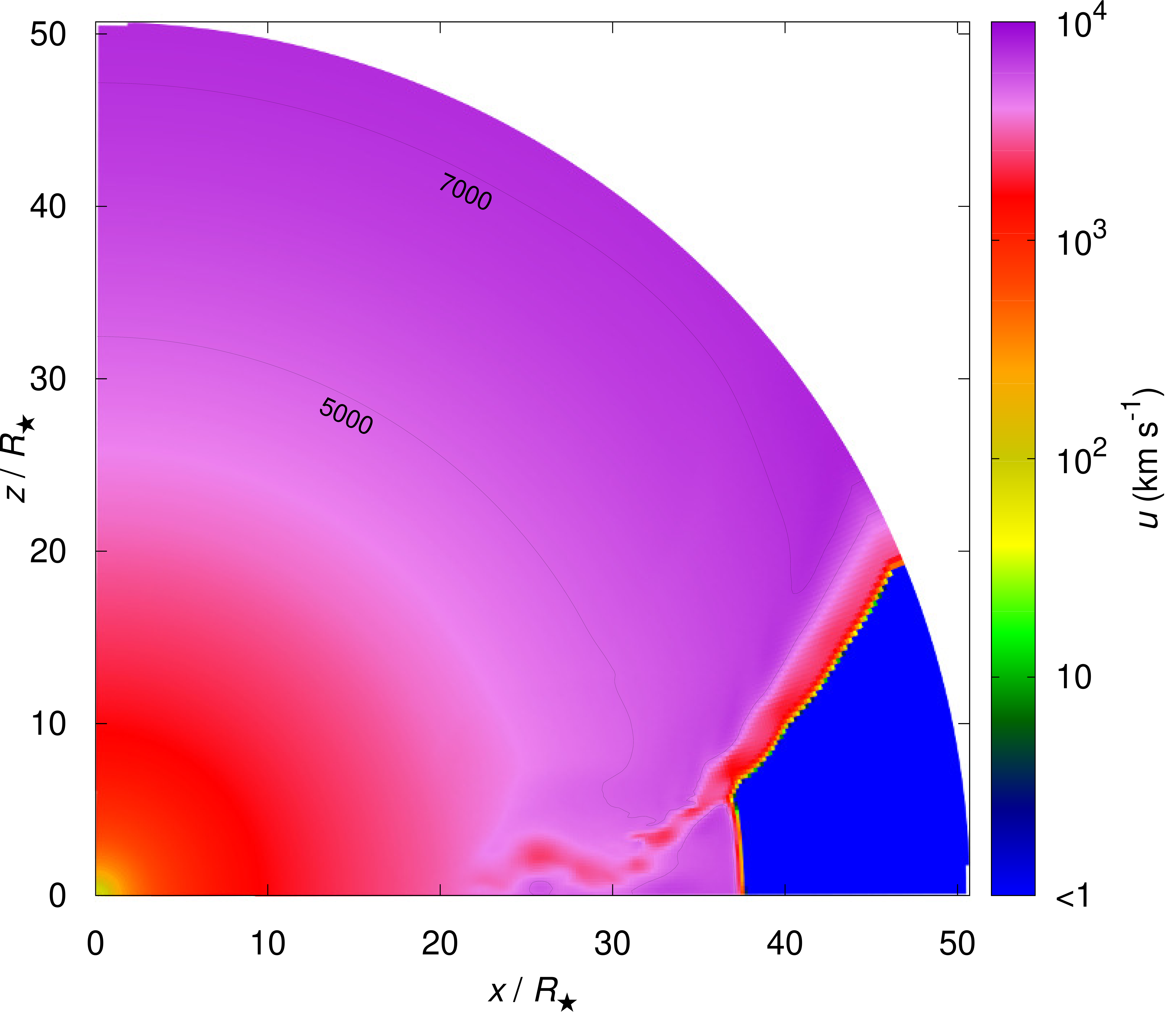}} 
\caption{Model C:
Detailed color maps of the density (\textit{upper panel}) and 
radial velocity (\textit{lower panel}) 
up to the distance $50\,R_\star$ at time $t \approx 105\,\text{h}$ 
since shock emergence, with the same resolution as in Fig. \ref{colorfigdensenew1}.
Contours denote the densities $\rho = 10^{-7},\,10^{-6},\,10^{-5},\text{ and }
10^{-4}\,\text{kg}\,\text{m}^{-3}$ and the velocities
$5000 \text{ and }7000\,\text{km\,s}^{-1}$.}
\label{colorfignew1}
\end{center}
\end{figure}
The asymetry of SN-CSM interaction zone is significant even in the case of
low density CSM (Model C).
The progenitor and CSM parameters are introduced in the first paragraph 
of Sect.~\ref{numero} and in Tab.~\ref{table1}.
The total masses of the spherical wind and circumstellar equatorial disk components
within the radius of the computational domain 
$r_1=20\,R_\star$ are $M_\text{sw}\approx 1.5\times 10^{-4}\,M_\odot$ and
$M_\text{csd}\approx 6.5\times 10^{-3}\,M_\odot$. 

We show (analogous to Sects.~\ref{numerodense} and \ref{numeromid}) in 
Figs.~\ref{colorfig21} and \ref{colorfig22} 
the 2D snapshots of density, velocity, and temperature, in equatorial ($x$) - polar ($z$) plane, 
at selected times $t=33\,\text{h}$ and $t=66\,\text{h}$ (only density). We again also add in 
Fig.~\ref{fig6_all}
the separated 
1D slices of graphs of density, velocity, pressure, temperature, and specific radiation entropy in equatorial
plane and in polar direction
at the time $t=33\,\text{h}$.
The figures in this case show the regions of significantly decelerated SN expansion velocity toward
the equatorial disk in greater distance
from the star than in the previous cases; in the corresponding time 
$t\approx 66\,\text{h}$ it is approximately $13\,R_\star$ (see Fig.~\ref{colorfig22}), 
while it is approximately $11\,R_\star$ in the previous models (cf.~Figs.~\ref{fig2_all} and \ref{fig4_all}). 
This is because the higher absolute value of expansion velocity even in this direction of 
maximum deceleration compared to previous models (cf. also Figs.~\ref{colorfig2}, 
\ref{colorfig11}, and \ref{fig6_all}).
The front of the polar SN expansion propagates with much higher 
velocity, i.e., $> 7000\,\text{km}\,\text{s}^{-1}$ while it is 
$\approx 4000\,\text{km}\,\text{s}^{-1}$ in the model with intermediate density
and $\approx 3000\,\text{km}\,\text{s}^{-1}$ in the model 
with high density in the same direction at the same time; the front should be within the same time $t\approx 66\,\text{h}$ 
obviously in much larger distance than in the previous models. This is why, outside the computational area, we plot
in this model the profiles of most of the characteristics in time $t\approx 33\,\text{h}$.

We show (analogous to Sects.~\ref{numerodense} and \ref{numeromid}) 
in the upper panel of Fig.~\ref{fig33_sparse} the density slopes $n$ and $w$ 
and in the lower panel the temperature slopes $p$,
calculated at the same time $t\approx 33\,\text{h}$ since shock emergence. 
The outer shock moves at the radius of approximately 
$9\,R_\star$ in the equatorial plane and 
$18\,R_\star$ in the polar direction.
The inner envelope density slope ranges in the polar direction between 
$n=0$ and $n\approx 17$; the temperature slope 
in the smooth region of the inner envelope 
ranges between $p=0$ and $p\approx 3.6$ in the equatorial and also in the polar direction.

The calculation of the applicability limits of the adiabatic self-similar solution (see Appendix \ref{limanal}) gives
for the model with low density the upper limit $t_\text{max}\approx 190\,\text{yrs}$ 
for the density slope $n=7,$ while 
for $n=12$ it gives $t_\text{max}\approx 3\times 10^4\,\text{yrs}$. 
The lower limit $t_\text{min}$ is still practically unimportant (cf. the model with high density
in Sect.~\ref{numerodense}).

As in both the previous models, we show in Fig.~\ref{colorfignew1} 
the snapshots of 2D density and expansion velocity in a more evolved stage (in time $t\approx 105\,\text{h}$
since shock emergence), up to the radius $50\,R_\star$, 
performed on 2D grid with 2400 grid cells in radial and 480 grid cells in azimuthal direction. 
Comparing this with Fig.~\ref{colorfig21}, we distinguish again the details of evolution of the selected 
quantities within the model.

\section{Conclusions and future work}\label{conclude}
We studied the hydrodynamic behavior of interaction between the  expanding envelope of SNe
and aspherical CSM containing an spherically symmetric stelar wind component and 
an aspherial circumstellar equatorial disk (or disk-like) component that is axially
symmetric and resides in the equatorial plane of the star. In the present model
we used simplified initial conditions, assuming homogeneous initial distribution 
of the stellar density and pressure without gravity ,
which has only a minor impact on the studied process of external interaction. 

We discussed three particular 2D models, supplemented by 1D illustrative slices 
of basic hydrodynamic quantities and of gradients (slopes) of density and temperature,
for three different values of stellar 
pre-explosion mass-loss rate that led to formation of CSM.
All the models significantly demonstrate the 
higher rate of deceleration of SN expansion in case of higher density of CSM and the
aspherical evolution of the density, velocity, and temperature structure of the SNe 
ejecta, where the mass preferably expands to the area outside the dense equatorial disk
\citep[cf.][]{1975ApJ...195..715M,1978ApJ...219L..23M,1981ApJ...247..908C}.
We conclude 
that in case the disk initial base density values 
$\rho_{0,\text{csd}}$ are of an order of magnitude (or more) higher than 
the base densities $\rho_{0,\text{sw}}$ of spherically symmetric CSM, 
such circumstellar equatorial disks effectively slow down
the equatorial SN expansion 
(compared to the SN expansion into regions outside the disk) 
with significant peaks of density, pressure, and temperature in the contact region of SN-disk interaction.
This is according to the presented models and other models with even lower initial
CSM densities
down to 
$\dot{M}_\text{csd}\approx 10^{-9}\,M_\odot\,\text{yr}^{-1}$ with $\rho_{0,\text{csd}}$
of an order of $10^{-9}\,\text{kg}\,\text{m}^{-3}$,
where the effects of asphericity vanish, and down to
$\dot{M}_\text{sw}\approx 10^{-7}\,M_\odot\,\text{yr}^{-1}$ with $\rho_{0,\text{sw}}$
of an order of 
$10^{-12}\text{ - }10^{-11}\,\text{kg}\,\text{m}^{-3}$ 
below which the effects of CSM become negligible. The comparison of power-law gradients of density and temperature 
(basically described in Figs.~\ref{fig33},~\ref{fig33_mid},~and~\ref{fig33_sparse}),
at least for the smooth parts of the expanding area (excluding the regions of steep shocks and discontinuities),
indicates their higher values for lower pre-explosion mass-loss rates. 
The slopes also decrease with time, however, this can be expected from
the similarity nature of expansion. We also checked the time limits of similarity solution applicability, 
where all the models match the time frame with considerable reserve.

The models also indicate the development of Kelvin-Helmholtz
instabilities in the zone of shear where the expanding matter flow is distorted along 
the interface between the spherically symmetric component and circumstellar equatorial disk component of CSM, 
forming the nearby layers of overdensity.

In our models, we did not consider the effects of radiative cooling, 
the influence of magnetic fields, or the energy input from the compact remnant of the explosion. 
These effects are typically mitigated and overlapped by the enormous SNe energy and their 
impact on the dynamics of the problem is very small \citep{1999ApJS..120..299T}, 
or are connected only with a specific situation or a type of the remnant 
\citep[like magnetars; see, e.g.,][]{2017ApJ...841...14M}.
The reverse shock formed in uniform ejecta is assumed to be radiative for a certain time \citep{1977ARA&A..15..175C},
but its velocity becomes eventually so high
and the ejecta are so tenuous that the cooling timescale for the gas
behind the shock exceeds the dynamical time of the SN expansion \citep{1999ApJS..120..299T}. 
It is therefore relevant to use the nonradiative approximation. In this point, \citet{1999ApJS..120..299T}
showed that the fraction of ejecta mass that cools is only a few percent in the case of $n<5$ ejecta,
in case of typical values of parameters. Moreover, \citet{1994ApJ...420..268C} 
studied the early radiative period for $n>5$ 
ejecta and found it had very little effect upon the dynamics of the SN expansion.

Because of the inclusion of radiative cooling, and therefore a violation of the energy conservation, the shocked shell
becomes thinner and denser than in the case of the used self-similar analytical description \citep{1982ApJ...258..790C}. 
Such a shell is however expected to be unstable because of developing (the Rayleigh-Taylor) instabilities 
\citep[e.g.,][]{1995ApJ...444..312C}, 
whose effect reduces the rate of conversion of the kinetic
energy to radiation \citep{1998ApJ...496..454B,2013MNRAS.428.1020M}.
These instabilities are usually included in 1D models
via an approximative ``smearing term'' 
(similar in principle to the numerical viscosity). We currently omit such
details
owing to the complexity of the SN-CSM interaction morphology, 
but they may be included in the future calculations 
\citep[cf. the study of][]{2018MNRAS.479..687S}.

Although the detailed description of observational signatures connected with the 
conclusions of the models is not the subject of this paper, we briefly comment on that issue. 
From an analytic solution suggested, for example, by 
\citet{2018ApJ...856...29M} (see also \citet{2013MNRAS.435.1520M}
or \citet{2013ApJ...773...76C}), it follows that in cases in which
the mass of the disk $M_\text{csd}$ is comparable with the mass of the ejecta $M_\text{ej}$, 
the ratio of the luminosity of the light curve peak is approximately indirectly proportional to the SN diffusion 
timescale $\tau_\text{sn}$; the time when the opacity drops so that the radiation can freely escape. In cases
in which the mass of the disk is
much smaller (two orders of magnitude or more) than the ejecta mass,
then the peak luminosity ratio of the light curve  is 
significantly reduced; the peak luminosities for the same ratio of the SN diffusion times are much closer.
Comparing the first two models A and B (where $M_\text{csd}$ is comparable with 
$M_\text{ej}\equiv M_\star$) with model C, 
we obtain a ratio of the diffusion timescales of about $3/2$. We may thus expect the light curve 
peak ratio to be approximately $1.2$ - $1.5$ in favor of model C. We may also assume that in case of yet lower 
disk masses the difference in luminosities would be even less distinct.
However, the situation may be complicated according to the line of sight of an observer. 
We expect that in the case of pole-on observational direction 
the difference in light curve luminosities would be
much smaller (if any) while in case of equator-on (disk-on) observational direction 
the effects of the disk may correspond to the above estimate.
We also expect that because of the asymetry of the outflow the resulting line
profiles should depend on the direction of observation showing higher outflow
velocities in the polar direction than in the equator-on direction.

As an immediate future step we plan to modify the initial stellar profiles of the density 
and temperature into a more realistic form, including 
the internal structure of the progenitor pre-calculated from stellar evolution code.
We also compare the current 2D models 
calculated without gravity with the models with the stellar gravity included. 
However, in this point we checked the influence of the 
gravitational force on the expansion profiles using the 1D SNEC calculations, and
there is a remarkable impact only in the central region of the 
original progenitor, which expands self-similarly together with the expansion of the whole envelope. For future 2D models we will implement the equilibrium-diffusion radiation transport solver into the code, 
taking into account recombination effects in the envelope and the effects of radioactive heating. This 
will enable us to 
perform the calculations in radiation-hydrodynamic mode with radiation-matter coupling.

We expect that subsequent study of the SN light curves powered by SN
thermal energy excess will
lead to a better understanding of the mass and density distribution of the CSM. 
It may also provide more precise estimates of the disk mass-loss rates and 
a deeper knowledge of the geometry and mechanism of the mass loss of massive stars.

\begin{acknowledgements}
We thank Dr.~Ond\v rej
Pejcha for helpful comments on details of the studied (or similar) process. We are also grateful to 
Milan Prv\' ak for technical support in preparation of the draft of this paper.
The access to computing and storage facilities owned by parties and projects contributing to the 
National Grid Infrastructure MetaCentrum, provided under the program 
``Projects of Large Infrastructure for Research, Development, and Innovations'' (LM2010005) is appreciated.
This work was supported by the grant GA \v{C}R
18-05665S and by the  grant Primus/SCI/17.
\end{acknowledgements}

\bibliographystyle{aa} 
\bibliography{bibliography} 

\begin{appendix}
\onecolumn
\section{Analytical solution of SN expansion}\label{analsol}

\subsection{Basics of {{adiabatic}} similarity solution}\label{basesimeqs}
Contact discontinuity separates the SN remnant
and CSM and propagates among the 
forward and reverse shock wave.
An analytical approach requires a similarity solution
(see, e.g., the formalism in \citet{1982ApJ...258..790C} or \citet{1985Ap&SS.112..225N}) 
with the coefficient of proportionality $\lambda(r,t)$ defined as $K\lambda(r,t)=rt^{-\alpha}$ 
(where the parameter $\alpha$ is different from $\alpha_{\text{vis}}$ in Sect.~\ref{baseq} 
and $K$ is the scaling constant). Denoting $\lambda_\text{I}=1$ the value of $\lambda$ 
at the inner (reverse) shock wave,
$\lambda_\text{C}$ at the contact discontinuity, and 
$\lambda_\text{II}$ at the outer forward shock wave [where obviously $\lambda_\text{in}
(r<r_\text{I})<\lambda_\text{I}<\lambda_\text{C}<\lambda_\text{II}<\lambda_\text{out}
(r>r_\text{II})$], coupling of the densities in Eq.~\eqref{homolog1} gives 
\begin{align}\label{homolog4}
\alpha=\frac{n-3}{n-w}.
\end{align}

Substituting the equations for $\lambda$ and $\alpha$ into Eq.~\eqref{homolog1}, 
we introduce the similarity variables $U(\lambda),V(\lambda),W(\lambda)$, and $C(\lambda)$, which 
correspond to velocity, density, pressure, and sound speed. The equations for $u,\rho,\text{ and }P$ 
are
\begin{gather}\nonumber
u(r\le r_\text{II})=Kt^{\alpha-1}\lambda U(\lambda),
\\u(r>r_\text{II})=0,\label{homolog5}\end{gather}
\begin{gather}\nonumber
\rho(r\le r_\text{C})=AK^{-n}t^{-\alpha w}V(\lambda),\\\rho(r>r_\text{C})=
BK^{-w}t^{-\alpha w}V(\lambda),\label{homolog5a}\end{gather}
\begin{gather}\nonumber
P(r\le r_\text{C})=AK^{2-n}t^{\alpha(2-w)-2}\lambda^2W(\lambda),\\P(r>r_\text{C})=
BK^{2-w}t^{\alpha(2-w)-2}\lambda^2W(\lambda).\label{homolog5b}
\end{gather}
The similarity variable $C(\lambda)$ is defined as $C^2=\gamma W(\lambda)/V(\lambda)$.
Following the relation for the coefficient $\lambda$, we may express the velocity $u_\text{C}$
of the contact discontinuity,
\begin{align}\label{aas4a}
u_\text{C}=\frac{\d r_\text{C}}{\d t}=\alpha\frac{r_\text{C}}{t},\quad U(\lambda_\text{C})=\alpha.
\end{align}
Substituting Eqs.~\eqref{homolog5} - \eqref{homolog5b} for the domain $r\le r_\text{II}$ 
into Eqs.~\eqref{mombase01} - \eqref{mombase21}, assuming spherical symmetry,
and partially differentiating with respect
to $t$ and $\lambda$, we obtain the following similarity relations 
\citep[cf.][]{1985Ap&SS.112..225N} for the continuity,
momentum, and energy equations, respectively,
\begin{align}\label{aaa0}
\lambda\left(U-\alpha\right)\frac{V^\prime}{V}+3U+\lambda U^\prime-\alpha w=0,
\end{align}
\begin{align}\label{aaa1}
\lambda\frac{V^\prime}{V}+\gamma\lambda U^\prime\frac{U-\alpha}{C^2}+
2\lambda\frac{C^\prime}{C}+\gamma U\frac{U-1}{C^2}+2=0,
\end{align}
\begin{gather}\label{aaa2}
\lambda\left(U-\alpha\right)\left[(1-\gamma)\frac{V^\prime}{V}+
2\frac{C^\prime}{C}\right]+2(U-1)+(\gamma-1)\alpha w=0.
\end{gather}
The explicit form of Eqs.~\eqref{aaa0} - \eqref{aaa2} for each particular similarity variable derivative is
\begin{align}\label{aaasim0}
\lambda U^\prime=-\frac{\mathcal{S}C^2+\left(U-\alpha\right)[3C^2-U(U-1)]}{C^2-\left(U-\alpha\right)^2},
\end{align}
\begin{align}\label{aaasim1}
\lambda\frac{V^\prime}{V}=\frac{\eta\mathcal{S}C^2+\left(U-\alpha\right)
[\left(U-\alpha\right)(3U-\alpha w)-U(U-1)]}
{\left(U-\alpha\right)[C^2-\left(U-\alpha\right)^2]},
\end{align}
\begin{align}
\lambda\frac{C^\prime}{C}=-\frac{\eta\mathcal{S}C^2}{2\left(U-\alpha\right)
[C^2-\left(U-\alpha\right)^2]}+
\frac{2C^2+(\gamma-1)U(U-1)+\left(U-\alpha\right)[2-(3\gamma-1)U]}
{2[C^2-\left(U-\alpha\right)^2]},\label{aaasim2}
\end{align} 
where we substitute the constant expressions $\eta$ and $\mathcal{S}$,
\begin{gather}\label{cano5}
\eta=\frac{(\gamma-1)\alpha w-2(1-\alpha)}{\alpha(3\gamma- w)-2(1-\alpha)}<1,
\end{gather}
\begin{gather}\label{cano4}
\mathcal{S}=\frac{\alpha(3\gamma- w)-2(1-\alpha)}{\gamma}>0.
\end{gather}
Multiplying Eq.~\eqref{aaa0} by an auxiliary constant $\beta$ and subtracting it from Eq.~\eqref{aaa2}, 
the integration gives the 
implicit
adiabaticity integral \citep[see, e.g.,][]{1959sdmm.book.....S,1985Ap&SS.112..225N} 
\begin{align}\label{adiabaticity}
\lambda^{2+3\beta}|U-\alpha|^\beta V^{1+\beta-\gamma}C^2=Q=\text{const.},
\end{align}
where the constant $\beta$ is
\begin{align}\label{lidov1a}
\beta=\frac{2(\alpha-1)+(\gamma-1)\alpha w}{\alpha(w-3)}.
\end{align}
 
The Rankine-Hugoniot relations \citep[e.g.,][]{1967pswh.book.....Z} give the $\lambda$-independent 
similarity variables for the inner strong shock,
\begin{gather}
U_{\text{I}}=\frac{2\alpha+\gamma-1}{\gamma+1},\quad V_{\text{I}}=\frac{\gamma+1}
{\gamma-1},\quad W_{\text{I}}=\frac{2(\alpha-1)^2}{\gamma+1},\quad 
C^2_{\text{I}}=\gamma\frac{W_{\text{I}}}{V_{\text{I}}}.\label{aaas5bb}
\end{gather}
The similarity variables for the outer strong shock are 
\begin{gather}
U_\text{II}=\frac{2\alpha}{\gamma+1},\quad 
W_\text{II}=\frac{2\alpha^2}{\gamma+1},\quad 
C^2_\text{II}=\gamma\frac{W_\text{II}}{V_\text{II}},\label{aaas5ab}
\end{gather}
where only the variable $V_\text{II}$ has to be evaluated numerically using the principles given in 
Appendix \ref{semianal}.

We also express the similarity variables in the zones inside the inner shock wave ($\lambda_\text{in}<1$) 
and outside the outer shock
wave ($\lambda_\text{out}>\lambda_{\text{II}}$). Combining Eq.~\eqref{homolog1} 
with Eqs.~\eqref{homolog5} - \eqref{homolog5b} and 
assuming $u_\text{out},\,P_\text{in},\,P_\text{out}=0$, we obtain
\begin{gather}
U_\text{in}=1,\quad V_\text{in}=\lambda_\text{in}^{-n},\quad C_\text{in}=0,\\
U_\text{out}=0,\quad V_\text{out}=\frac{\gamma-1}{\gamma+1}\left(\frac{\lambda_\text{II}}
{\lambda_\text{out}}\right)^{w}V_\text{II},\quad C_\text{out}=0.\label{aoutr}
\end{gather}

From Eq.~\eqref{aaasim0} it follows that in very proximity of the contact discontinuity (where $U-\alpha\rightarrow 0$) 
$\lambda U^\prime=-\mathcal{S}$. Integrating the left-hand side
of this relation from $U(\lambda_\text{C})$ 
to a nearby $U$ and the right-hand side from $\lambda_\text{C}$ to a nearby $\lambda$, as well as integrating analogously
Eqs.~\eqref{aaasim1} - \eqref{aaasim2}, gives the similarity relations close to $r_\text{C}$
in the form
\begin{align}\label{conias1}
U-\alpha=-\mathcal{S}\ln\frac{\lambda}{\lambda_\text{C}},\quad V=\xi\left|U-\alpha\right|^{-\eta},\quad C^2=\zeta\left|U-\alpha\right|^{\eta},
\end{align}
where $\xi$ and $\zeta$ are arbitrary constants that differ in general on both sides of $r_\text{C}$. However, since $u$ and $P$ 
are continuous \citep{1967pswh.book.....Z} through $r_\text{C}$  
(implying a constant product $VC^2$ in Eq.~\eqref{conias1}), 
we may write $(\xi\zeta)_\text{i}=(\xi\zeta)_\text{o}$,
where the subscript ``i'' denotes the zone between the inner shock wave and the contact discontinuity 
while the subscript ``o'' denotes the zone between the contact discontinuity and the outer shock wave.
To determine the constants $\xi$ and $\zeta$, Eqs.~\eqref{conias1} have to be calculated numerically within the 
semi-analytical solution (see Sect.~\ref{semianal}).

We obtain the relations for total mass $M_\text{I,II}$ and total energy $E_\text{I,II}$ 
confined between the two shocks \citep{1985Ap&SS.112..225N} by integration of Eq.~\eqref{homolog1},
with use of the formalism in this Section,
\begin{gather}\label{aaas7a} 
M_\text{I,C}=\frac{4\pi A}{(n-3)K^{n-3}}t^{\,\alpha(3-w)},\quad 
M_\text{C,II}=\frac{\gamma-1}{\gamma+1}\frac{4\pi BK^{3-w}}{3-w}
V_\text{II}\lambda_\text{II}^3\,t^{\,\alpha(3-w)},\quad 
E_\text{I,II}=\frac{2\pi A}{(n-5)K^{n-5}}t^{(1-\alpha)(n-5)}.
\end{gather}
This leads to the important constraint $w<3$ and $n>5$ for the similarity solution, while 
Eq.~\eqref{homolog4} implies the condition $\alpha<1$. 

  To evaluate the constants $A,\,B,\text{ and }K$, we identify the mass 
  $M_\text{ej}$ of the SN envelope as $M_\text{I,C}$ (which tends to be relevant in an advanced time) 
  and the energy $E_\text{ej}$ of the envelope as $E_\text{I,II}$.
  Integrating the first equation
  \eqref{homolog5a} from $r_\text{I}$ to infinity, with use of Eq.~\eqref{aaas7a}, gives the constant $A$.
  The constant $B$ results from the mass conservation equation for CSM (cf.~Eq.~\eqref{homolog1}) 
  as $B=\dot{M}_\text{sw}/(4\pi R_\star^{2-w}\varv_\text{sw})$,
  assuming the stationary pre-explosion CSM velocity $\varv_\text{sw}\ll u(r<r_\text{II})$.
  The constant $K$ results from Eq.~\eqref{homolog5b} following the continuous pressure at the 
  contact discontinuity $r_\text{C}$. 
  This also implies $BK^{j-w}\equiv A/K^{n-j}$, where $j$ is an arbitrary integer.

\subsection{Principles of semi-analytical solution of the similarity equations}
\label{semianal}
We basically adopt the principles of the numerical integration of the analytically expressed 
solution given in Sect.~\ref{sedov} 
from \citet{1985Ap&SS.112..225N}.
We first numerically integrate Eqs.~\eqref{aaasim0} - \eqref{aaasim2} with use of 
constant adiabaticity integral, Eq.~\eqref{adiabaticity}, 
in the region ``i'' (where $r_\text{I}<r<r_\text{C}$), until the condition $0<U-\alpha\ll\alpha$ is satisfied.
The input values of the similarity variables for evaluation of the constant $Q=Q_\text{i}$ in 
\eqref{adiabaticity} are given by Eq.~\eqref{aaas5bb}.
We put the similarity variables $U,V,C,\text{ and }\lambda$, calculated in the previous step, 
to Eqs.~\eqref{conias1}. 
We thus determine $\xi_\text{i}$, $\zeta_\text{i}$, and $\lambda_\text{C}$ to complete 
the integration between $r_\text{I}$ and $r_\text{C}$. 
Then we integrate the 
region ``o'' (where $r_\text{C}<r<r_\text{II}$) in a similar way by inserting an input value 
of $\lambda$ slightly larger than $\lambda_\text{C}$,  
until the condition $0<U-\alpha\ll\alpha$ is satisfied (Eq.~\eqref{conias1}) again. 

We select a trial value of the constant $\xi_\text{o}$ to determine 
the constant $\zeta_\text{o}$. The analogous initial conditions for the constant 
$Q=Q_\text{o}$ in \eqref{adiabaticity} ,
noting that $Q_\text{o}\ne Q_\text{i}$ in general, 
are given by Eq.~\eqref{aaas5ab}.
Integration continues until we reach the equality given by the second equation 
\eqref{homolog5b} for $\lambda=\lambda_\text{II}$. 
Because the value of $C_\text{II}$ differs in this case from the value given by Eq.~\eqref{aaas5ab}, 
we use an algorithm that finds such a constant $\zeta_\text{o}$, for which Eq.~\eqref{aaas5ab} is satisfied with 
the required accuracy while the second equation 
\eqref{homolog5b} holds as well. Once this $\zeta_\text{o}$ value is found, the problem is completely resolved 
because in this case the values of the similarity variables $V_\text{II}$ and $C_\text{II}$ are also uniquely determined.
We study the cases with $\gamma_\text{o}=\gamma_\text{i}$ 
(see explanation in Sect.~\ref{sedov}) and $\mu_\text{o}=\mu_\text{i}$;
the possible difference between the mean molecular weight $\mu$ of the matter 
expelled from the star and that of the ambient medium has no 
effect on the structure of the similarity solution \citep{1985Ap&SS.112..225N}. 

\subsection{Constraints of the adiabatic self-similar solution}
\label{limanal}
Regarding the early stages of SN expansion, we may neglect the energy losses due to the volume radiation 
\citep{1985Ap&SS.112..225N}.
However, the similarity solution is yet limited by the applicability of the adiabatic gas dynamics. The first constraint
results from the condition that the 
length of the mean free path of the particles must be small compared to the characteristic scale $\Delta r$ 
of the expanding envelope. In other words, the width $\ell$ of the shock front 
must be smaller than  the characteristic scale, $\ell< \Delta r$. 
Following the study of a strong shock wave front in hydrogen plasma
by \citet{1975FizPl...1..202I} (cf. also \citet{1985Ap&SS.112..225N}), we adopt the value (in SI units)
\begin{align}\label{lim1}
\ell\simeq \frac{8.1\times 10^{22}}{n_\text{in}}
\left(\frac{\widetilde{u}_\text{in}}{10^3\,\text{km}\,\text{s}^{-1}}\right)^4\,\text{m},
\end{align}
where $n_\text{in}=\rho_\text{in}/m_\text{H}$ is the number density of hydrogen atoms before the shock wave front 
($m_\text{H}$ is the mass of a hydrogen atom) and $\widetilde{u}_\text{in}$
is the relative flow velocity of the expanding gas before the corresponding 
shock wave (in the rest frame of the shock wave).
Taking the distance $r_\text{I,C}=|r_\text{I}-r_\text{C}|$ between the inner shock wave and the contact discontinuity as 
the characteristic scale $\Delta r$ , where we do not consider the outer shock wave because it propagates outside 
the SN gas, $r_\text{II}>r_\text{C}$, 
following the definition of the coefficient $\lambda$ gives
\begin{align}\label{lim2}
\ell<(\lambda_\text{C}-1)\,Kt^\alpha.
\end{align}
We evaluate the velocity $\widetilde{u}_\text{in}$ as $u_\text{I}-u_\text{in}$ 
following the strong shock Rankine-Hugoniot equations \citep[e.g.,][]{1967pswh.book.....Z}.
Employing Eqs.~\eqref{homolog5} and \eqref{aaas5bb} gives
\begin{align}\label{lim3}
\widetilde{u}_\text{in}=(1-\alpha)\frac{r_\text{I}}{t}.
\end{align}

The second constraint obviously results from the finite amount of the ejected matter 
$M_\text{ej}$. The total mass in the SN envelope must obey the
inequality
\begin{align}\label{lim4}
M_\text{I,C}<M_\text{ej},
\end{align}
where $M_\text{I,C}$ is given by Eq.~\eqref{aaas7a}.
The adiabatic similarity solution is thus applicable within the time interval 
$t_\text{min}<t<t_\text{max}$, where we obtain $t_\text{min}$ from Eq.~\eqref{lim2} and 
$t_\text{max}$ results from Eq.~\eqref{lim4}. 

\end{appendix}

\end{document}